# Analyzing the Carrier Mobility in Transition-metal Dichalcogenide MoS$_2$ Field-effect Transistors

*Zhihao Yu, Zhun-Yong Ong, Songlin Li, Jian-Bin Xu, Gang Zhang, Yong-Wei Zhang, Yi Shi and Xinran Wang*

Z. Yu, Prof. S. Li, Prof. Y. Shi, Prof. X. Wang
National Laboratory of Solid State Microstructures, School of Electronic Science and Engineering, and Collaborative Innovation Center of Advanced Microstructures
Nanjing University, Nanjing 210093, P. R. China
E-mail: xrwang@nju.edu.cn, yshi@nju.edu.cn
Dr. Z. Y. Ong, Dr. G. Zhang, Dr. Y. W. Zhang
Institute of High Performance Computing, 1 Fusionopolis Way, 138632, Singapore
E-mail: zhangg@ihpc.a-star.edu.sg
Prof. J. Xu
Department of Electronic Engineering and Materials Science and Technology Research Center
The Chinese University of Hong Kong, Hong Kong SAR, P. R. China

**ABSTRACT**

Transition-metal dichalcogenides (TMDCs) are important class of two-dimensional (2D) layered materials for electronic and optoelectronic applications, due to their ultimate body thickness, sizable and tunable bandgap, and decent theoretical room-temperature mobility of hundreds to thousands cm$^2$/Vs. So far, however, all TMDCs show much lower mobility experimentally because of the collective effects by foreign impurities, which has become one of the most important limitations for their device applications. Here, taking MoS$_2$ as an example, we review the key factors that bring down the mobility in TMDC transistors, including phonons, charged impurities, defects, and charge traps. We introduce a theoretical model that quantitatively captures the scaling of mobility with temperature, carrier density and thickness. By fitting the available mobility data from literature over the past few years, we are able to obtain the density of impurities and traps for a wide range of transistor structures. We show that interface engineering such as oxide surface passivation, high-k dielectrics and BN encapsulation could effectively reduce the impurities, leading to improved device performances. For few-layer TMDCs, we analytically model the lopsided carrier distribution to elucidate the experimental increase of mobility with the number of layers. From our analysis, it is clear that the charge transport in TMDC samples is a very complex problem that must be handled carefully. We hope that this Review can provide new insights and serve as a starting point for further improving the performance of TMDC transistors.

# 1. Introduction

As the dimension of Si metal-oxide-semiconductor field-effect transistors (MOSFETs) continues to scale down, the speed and integration level of semiconductor chips have progressed by leaps and bounds in the past 60 years.[1] Moore's law has led the semiconductor industry across half a century to the 14nm technology node through continuous process innovations.[2, 3] To maintain the progress of Moore's law, the quasi-2D FinFET structure has been adopted after the 32nm node.[4] However, as the process complexity increasing and the ultimate scale approaching, the pernicious performance such as larger leakage current and subthreshold slope (SS) induced by shot-channel effects cannot be suppressed through new processes alone.[5] The International Technology Roadmap for Semiconductors (ITRS)[6] provides guidelines on new generation complementary metal–oxide–semiconductor (CMOS) technology, showing that new materials beyond silicon will be needed to continue the scaling beyond the 7nm technology node[7], the so-called "More Moore" approach.

In fact, the materials-based physical limitation of channel scaling can be calculated from the natural length scale of MOSFETs as $\lambda = \sqrt{\frac{\varepsilon_{ch}}{N\varepsilon_{ox}} d_{ch} d_{ox}}$, where $\varepsilon_{ch}$ and $\varepsilon_{ox}$ is the dielectric constant of channel and gate oxide, $d_{ch}$ and $d_{ox}$ is the thickness of channel and gate oxide[8]. To achieve effective gate control, the channel length needs to exceed $4\lambda$. It is clear that one can continue the scaling by reducing the thickness of the channel. Given the quantum limitation of bulk semiconductors like silicon, the channel thickness is hardly thinner than 5nm.[6] The mobility of bulk semiconductors will also dramatically decrease as the channel thickness decreases, due to unavoidable surface roughness scattering.[9] Thus, this is a great challenge that all post-silicon electronics has to face. 2D semiconductors (2DSCs) are ideal candidates to further advance CMOS technology to smaller device dimensions.[10] Like graphene, they are layered in nature and have the ultimate thickness down to one atomic layer.[11] Unlike graphene, which is a zero-gap material, 2DSCs have a sizable bandgap up to several electronvolts,[12, 13] making them suitable for applications like logic transistors[14-18] flexible electronics[19-23] and photodetectors[24-28]. It is anticipated that monolayer 2DSCs can scale the transistors down to 3 nm gate length. The electron velocity of 2DSCs reaches $5\times10^6$~$5\times10^7$cm/s in ballistic region, compared with silicon($1\times10^7$cm/s)[29]. Moreover, the possible large-area material synthesis by chemical vapor deposition (CVD) shows great promise in wafer-scale device integration [30-34].

So far, tens of 2DSCs have been studied with very different properties.[35-37] Some of the prominent examples include TMDCs, monochalcogenides,[38-41] and graphene-like X-ene (e.g. black phosphorus and silicene)[14, 42, 43]. Among these materials, TMDCs are most promising to integrate with CMOS under the "More Moore" scheme, considering their carrier mobility, bandgap, material stability and the possibility of high-quality, large-area synthesis. [44] TMDCs have a general formula of $MX_2$, where M is transition-metal and X is chalcogen. A monolayer of TMDCs has a X-M-X trilayer sandwich structure, where the M and X atoms stack with either trigonal prismatic (2H phase)[45] or octahedral coordination (1T phase)[46] geometries. The idea of using TMDCs for transistor applications had been proposed around 2004-2005.[47, 48] However, their potential was not fully acknowledged until *Kis* et al. reported high-performance monolayer 2H-MoS$_2$ (abbreviated as MoS$_2$ in the rest of this Review) topgate transistors in 2011.[49] Since then, enormous progress has been made to understand the charge transport[50], interfacial properties[51] and electrical contacts[52-54] of MoS$_2$ and other semiconducting group-VI TMDCs[55], whose bandgaps are ~1-2 eV. Controlled doping[56-59] and device integration[60-63] are also of great interest. Recently, the research on TMDCs has expanded to other groups of transition metals, which extend the bandgap spectrum to mid-infrared.[64] Figure 1a shows the position of TMDCs on the mobility-bandgap chart, along with other important electronic materials. The carrier mobility on the order of 100-1000cm$^2$/Vs even in the monolayer limit, tunable bandgap up to ~2eV, and large family of complementary materials make TMDCs stand out as one of the most promising materials. Moreover, new quantum phenomena in TMDCs such as charge-density waves [67], topological transport[68] and valleytronics[69] have been proposed to be implemented in "beyond CMOS" device applications. We note that Figure 1a displays the bulk properties of conventional Si, Ge and III-V semiconductors. It is well know that the mobility of bulk semiconductors will be strongly degraded by surface roughness scattering at the 2D limit, making TMDCs even more favourable. For example, when the thickness of Si body ($t_s$) is less than 4 nm, the mobility is roughly proportional to $t_s^6$, which is attributed to the surface-roughness scattering induced by device processing.[65-66]

For future heterogeneous integration, the processing of new materials must be fully compatible with CMOS technology. Recently, several breakthroughs have been made in this direction. In 2014, a 3D FET with MoS$_2$ channel covering HfO$_2$/Si upstanding Fin structure was demostrated. Low operating voltage down to 1V and matched $V_{th}$ between the n-type and p-type transport regime were realized in the hybrid Si/TMDC FETs with 50nm channel length and equivalent oxide thickness (EOT) <1nm.[70] More recently, *Chen* et al. demonstrated an ultra-scaled U-shape MoS$_2$ transistor with self-aligned Si source/drain and Si topgate (Figure 1b, c), by a fully CMOS-compatible process. They used low temperature CVD (755℃) to deposit few-layer MoS$_2$ in the pre-defined channel region with full-wafer availability.[71] Cross-sectional transmission electron microscopy (TEM) image in Figure 1c shows that the MoS$_2$ channel sits perfectly at the edge of source/drain electrodes, encapsulated by the gate electrode. Even without careful optimizations, these 10nm-channel MoS$_2$ transistor already showed very respectful performance: on/off ratio of $10^5$ and on-state current of 150 $\mu A/\mu m$. These works clearly demonstrate the prospects of MoS$_2$ in extending the Moore's law beyond Si. To the best of our knowledge, no other 2DSCs have been succesfully integrated with Si at similar level yet.

In spite of the great promise as a transistor channel material, it has become clear that the electron transpot of MoS$_2$ and other TMDCs is still severely limited by extrinsic factors. This is mainly manifested by the much lower experimental mobility than the theoretically predicted phonon-limited values, even in the state-of-the-art devices. For example, the theoretical phonon-limited mobilty of monolayer $MoS_2$ and $WS_2$ at room temperature is ~410cm$^2$/Vs[72] and ~1100cm$^2$/Vs [73]. However, the experimental record is only 150cm$^2$/Vs and 80cm$^2$/Vs for these two materials, respectively[74, 75]. This is one of the key limitations in the device applications of TMDCs that deserves keen attention. Among the major extrinsic factors are Coulomb impurities (CIs) near the semiconductor-dielectric interface, charge traps, defects, surface optical (SO) phonons and thickness, which all play important roles in scattering charge carriers. It is thus paramount to theoretically understand each scattering mechanism and their signatures in electron transport. Furthermore, a model including the major sources of scattering should be developed to quantitatively diagnose the transistor data, find the underlying bottlenecks, and provide guidelines for further improvement. Over the last few years, we have built such a model that works extremely well in our own $MoS_2$ and $WS_2$ devices[74-76] In this review, taking $MoS_2$ as an example, we use this model to analyze the available field-effect mobility data in the literature and extract important microscopic quantities such as the density of CIs and charge traps in various transistor structures. We further generalize the model to few-layer $MoS_2$ FETs, where the finite width of charge distribution has to be considered. We note that there already exist several outstanding reviews of the electronic properties[12, 50-53], synthesis[77-79] and device applications of TMDCs[35, 38, 80-83]. Many of them are devoted to equally important topics such as contact resistance[52-53], large-scale material synthesis[84-86], and heterostructures[73,84]. The purpose of this review is not to seek comprehensive coverage of similar topics, but rather to dig deeper into the transistor data, understand the underlying device physics, and compare various transistor structures. We hope that our results provide useful insights to the readers and a clear path to realizing full potential of TMDCs in high-performance FETs. Although the review is mainly focused on $MoS_2$ due to limtied space, we believe the conclusions are generic for all related TMDCs.

The review is organized as following. In the second section, we review the scattering mechnisms in $MoS_2$ from theoretical point of view, and establish our theoretical model with detailed parameters. The experimental evidence of short-range defects and charge traps is also presented to rationalize our model. In the third section, we use our model to analyze the available monolayer $MoS_2$ FET data from literature. We shed light on the microscopic origin behind the low mobility in various $MoS_2$ FET structures. In the fourth section, we discuss experiment progress and present a model for few-layer $MoS_2$ FETs, taking the finite thickness into account. Finally, we draw some conclusions and future perpectives from our analysis.

**2. Electron scattering mechanisms in monolayer $MoS_2$**

So far, with only a few exceptions [87-88], most of the reported $MoS_2$ FETs operate in the diffusive regime, where the channel length is much longer than the electron mean free path. Typically, the charge transport in monolayer $MoS_2$ can be divided into the insulating and metallic regimes, depending on the carrier density. At sufficiently high carrier density, the system transits into the metallic regime, allowing us to treat the system as a 2D electron gas. Thus, the electron mobility in MoS$_2$ can be described using a semi-classical transport model based on the Boltzmann transport equation (BTE), the relaxation time approximation and Matthiesen's rule [89]. The model we have describes the mobility which is related to the charge carrier diffusivity. The use of the concept of mobility implicitly assumes that the device is large enough so that the movement of charge is diffusive in nature, not ballistic. The electrons are assumed to move diffusively with an effective mass [72] of $m^* = 0.48m_0$, where $m_0$ is the free electron mass. The expression for each components of the electron mobility is given by [89-90]

$$\mu = \frac{2e}{\pi n \hbar^2 k_B T} \int_0^\infty f(E)[1-f(E)]\Gamma(E)^{-1} E dE \qquad (1)$$

where $e$, $n$, $\hbar$, $k_B$ and $T$ are the electron charge quantum, the electron density, the Planck constant divided by $2\pi$, the Boltzmann constant and the temperature, respectively. The functions $f(E)$ and $\Gamma(E)$ in turn represent the Fermi-Dirac distribution and the momentum relaxation rate due to electron scattering by intrinsic and extrinsic processes which we will discuss later. The electron band mobility is estimated using Matthiesen's rule [89]:

$$\mu_0(n,T)^{-1} = [\mu_{SO}(T)^{-1} + \mu_{ph}(T)^{-1}] + \mu_{CI}(n,T)^{-1} \qquad (2)$$

where $\mu_{ph}$, $\mu_{SO}$ and $\mu_{CI}$ are the individual mobility components due to scattering with intrinsic phonons, SO phonons, and CIs, respectively, and are calculated using Eq. (1) with the respective scattering rates. The details of how the scattering rates and the mobility components are calculated are discussed in the following sections and based on the simplified modeling method used in Ref. 75. In particular, we pay attention to how they affect the temperature- and carrier-density-dependence of the mobility.

**2.1 Intrinsic electron-phonon scattering**

In the absence of any external scattering sources, the electron mobility within the MoS$_2$ monolayer is limited by the interaction between electrons and the various intrinsic types of lattice phonons. This intrinsic phonon-limited mobility is important as it determines the maximum electron mobility that can be achieved in very clean samples that are free of impurities and defects. It also sets the benchmark to which experimental mobility values can be compared for characterizing the quality of the sample.

*Kaasbjerg* et al. [72] identified the intrinsic phonon scattering rate ($\Gamma_{ph}$) as originating from the scattering of electrons by the longitudinal (LA) and transverse acoustic (TA), the intravalley polar longitudinal optical (Froehlich), the intervalley polar longitudinal optical (LO) and the intravalley homopolar optical (HP) phonons. In the context of analyzing the electron mobility, these phonon types differ mainly in terms of their relative importance at different temperature regimes and to a much lesser extent, how they vary with electron density. The momentum relaxation rate associated with intrinsic phonon scattering is $\Gamma_{ph} = \Gamma_{LA} + \Gamma_{TA} + \Gamma_{LO} + \Gamma_{HP} + \Gamma_{Fr}$, where $\Gamma_{LA}$, $\Gamma_{TA}$, $\Gamma_{LO}$, $\Gamma_{HP}$ and $\Gamma_{Fr}$ are the scattering rates associated with LA, TA, LO, HP and Froehlich phonons, respectively.

At low temperatures ($T < 100$ K), the intrinsic phonon-limited mobility is dominated by scattering with the low-frequency, long-wavelength LA and TA phonons like in conventional semiconductors. The LA and TA phonon scattering rates are given by $\Gamma_{ac} = \frac{m^* \Xi_{ac}^2 k_B T}{\hbar^3 \rho c_{ac}^2}$, where $\Xi_{ac}$ is the acoustic phonon (LA or TA) deformation potential and $c_{ac}$ is the acoustic phonon speed. The linear dependence of the scattering rates on $T$ implies that the low-temperature intrinsic phonon-limited mobility scales as $T^{-1}$.

On the other hand, the effects of optical phonon (LO, HP and Froehlich) scattering on the mobility become more pronounced at higher temperatures as these phonon states become more thermally populated. The zero-th order intervalley polar longitudinal optical (LO) and intravalley homopolar optical (HP) phonon scattering rates can be calculated using the formula [91] $\Gamma_{op}(E) = \frac{m^* D_{op}^2}{2\hbar^2 \rho \omega_{op}}[N_{op} + (1+N_{op})\Theta(E - \hbar\omega_{op})]$, where $D_{op}$ is the optical deformation potential of the optical phonon (LO or HP) and $N_{op} = [\exp(\hbar\omega_{op}/k_V T) - 1]^{-1}$ is its Bose-Einstein distribution with $\omega_{op}$ its phonon energy. $\Theta(...)$ is the usual Heaviside function. Higher-order corrections to the scattering rates are possible although numerically, they do not significantly affect the overall intrinsic phonon-limited mobility.

Unlike the intervalley polar LO and intravalley HP phonons, electron interaction with the *intra*valley polar longitudinal optical (Froehlich) phonons is probably dependent on the electron density because the bare electron-phonon coupling is charge-dependent and in the long wavelength limit, the intravalley polar longitudinal optical (Froehlich) phonons can couple with the electron gas and undergo screening to reduce the effective charge, i.e. $e/\varepsilon_{ion}^0 \to e/[\varepsilon_{ion}^0 + \varepsilon_{el}(q)]$. Thus, we can write the Froehlich optical phonon emission (+) or absorption (-) rate as [91]:

$$\Gamma_{Fr}^{\pm}(E) = \frac{e^2 \omega_{Fr} m^*}{8\pi\hbar^2}\left(\frac{1}{2} \pm \frac{1}{2} + N_{Fr}\right)\int_{-\pi}^{\pi} d\theta \frac{1-\frac{k'}{k}\cos\theta}{q}\left(\frac{1}{\varepsilon_{ion}^{\infty}+\varepsilon_{el}(q)} - \frac{1}{\varepsilon_{ion}^0+\varepsilon_{el}(q)}\right)\text{erfc}(\frac{q\sigma}{2})^2$$

where $k$ is the initial state, $k'$ is the final state given by $k'=\sqrt{k^2 \mp 2m^*\omega_{Fr}/\hbar}$, and $q=\sqrt{k^2+k'^2-2kk'\cos\theta}$; $\varepsilon_{ion}^{\infty}$ ($\varepsilon_{ion}^0$) is the ionic part of the optical (static) permittivity of MoS$_2$, erfc is the complementary error function and $\sigma$ is the sheet thickness. $\varepsilon_{el}(q) = -\frac{e^2}{2q}\Pi(q,T,E_F)$ corresponds to the electronic part of the dielectric function and $\Pi(q,T,E_F)$ is the temperature- and carrier density-dependent static polarizability, and represents the polarization charge screening of the CI. Its exact form is given in Refs. [90],[91],[92]. We assume the phonon dispersion for the LO phonons to be flat so that $\omega_{Fr} = \omega_{LO}$. Therefore, the Froehlich phonon scattering rate is: $\Gamma_{Fr}(E) = \Gamma_{Fr}^{-}(E) + \Theta(E-\hbar\omega_{Fr})\Gamma_{Fr}^{+}(E)$.

Figure 2(a) shows the intrinsic phonon-limited mobility $\mu_{ph}$ as a function of temperature at $n = 10^{12}$ and $10^{13}$ cm$^{-2}$, calculated using Eq. (1) and the parameters in Table 1. At low temperatures, the intrinsic phonon-limited mobility scales as $\mu \propto T^{-1}$ because of acoustic phonon scattering. However, scattering with optical phonons takes over as the dominant scattering mechanism at higher temperatures and the mobility decreases more rapidly with temperature, scaling as $\mu \propto T^{-1.7}$. This mobility temperature scaling at room temperature is sometimes used to identify the dominant mechanism limiting charge transport in the device [93,94], although caution should be exercised. At 300 K, we find that $\mu_{ph}$ ~410 cm$^2$/Vs which is the maximum room-temperature electron mobility if there is no scattering with Coulomb impurities and surface optical phonons and in good agreement with published theoretical results [72, 91]. There is however no significant carrier density dependence, which indicates that screening in the Froehlich interaction is not a major factor in the intrinsic phonon-limited mobility.

**2.2 SO phonon scattering**

Apart from electron scattering with the intrinsic phonons, another important source of inelastic electron scattering is through remote interaction with the polar optical phonons in the dielectric substrate [69, 91] and the top gate oxide in the dual-gated device. In an oxide insulator like SiO$_2$, the metal-oxide bonds are easily polarizable and the oscillatory motion of these bonds from the *polar* optical phonon modes produces a time-dependent evanescent field at the substrate surface that can scatter electrons in the MoS$_2$. In commonly used oxide insulators such as SiO$_2$, Al$_2$O$_3$ and HfO$_2$, inelastic electron scattering by the polar surface optical (SO) phonons can significantly reduce the electron mobility. The amount of scattering depends on the dielectric coupling strength and the characteristic frequency of these phonons. In particular, low-frequency phonon modes have a pronounced deleterious effect on the electron mobility because they are significantly populated at room temperature. It is usually assumed in the modeling of electron scattering by SO phonons that there are two SO phonon modes in oxides like SiO$_2$ and Al$_2$O$_3$, and the SO phonon scattering rate is given by the sum of their individual scattering rates, *i.e.*, $\Gamma_{so} = \Gamma_{so1} + \Gamma_{so2}$.

For HfO$_2$ though, we can assume that there is only one phonon mode like in Ref. [95]. For h-BN, there are also two SO phonon modes ($\omega_{TO1}$ = 97.4 meV and $\omega_{TO2}$ = 187.9 meV) although the higher one has virtually no effect on the carrier mobility. The dielectric function of the substrate $\varepsilon_{box}(\omega)$ describes the frequency response of the dielectric and is[90] [96]:

$$\varepsilon_{box}(\omega) = \varepsilon_{box}^{\infty} + (\varepsilon_{box}^{i} - \varepsilon_{box}^{\infty})\frac{\omega_{TO2}^2}{\omega_{TO2}^2 - \omega^2} + (\varepsilon_{box}^{0} - \varepsilon_{box}^{i})\frac{\omega_{TO1}^2}{\omega_{TO1}^2 - \omega^2}$$

where $\varepsilon_{box}^{0}$, $\varepsilon_{box}^{i}$ and $\varepsilon_{box}^{\infty}$ are the static, intermediate and optical dielectric of the substrate, respectively, and $\omega_{TO1}$ and $\omega_{TO2}$ are the transverse optical (TO) phonon angular frequencies such that $\omega_{TO1} < \omega_{TO2}$. We can rewrite $\varepsilon_{box}(\omega)$ in the generalized Lyddane-Sachs-Teller form $\varepsilon_{box}(\omega) = \varepsilon_{box}^{\infty}\left(\frac{\omega_{LO1}^2 - \omega^2}{\omega_{TO1}^2 - \omega^2}\right)\left(\frac{\omega_{LO2}^2 - \omega^2}{\omega_{TO2}^2 - \omega^2}\right)$. In the case of HfO$_2$ where there is only one TO mode, its dielectric function is

$$\varepsilon_{box}(\omega) = \varepsilon_{box}^{\infty} + (\varepsilon_{box}^{0} - \varepsilon_{box}^{\infty})\frac{\omega_{TO}^2}{\omega_{TO}^2 - \omega^2} = \varepsilon_{box}^{\infty}\left(\frac{\omega_{LO}^2 - \omega^2}{\omega_{TO}^2 - \omega^2}\right)$$

while $\varepsilon_{tot,SO}^{\infty} = \frac{1}{2}\left(\varepsilon_{box}^{\infty} + \varepsilon_{0}\right)$ and $\varepsilon_{tot,SO}^{0} = \frac{1}{2}\left[\varepsilon_{box}^{\infty}\left(\frac{\omega_{LO}^2}{\omega_{TO}^2}\right) + \varepsilon_{0}\right]$. The SO phonon frequencies ($\omega_{SO1}$ and $\omega_{SO2}$) are determined from the roots of the equation $\varepsilon_{box}(\omega) + \varepsilon_{0} = 0$.

The remote interaction between the electrons and the substrate SO phonons depends on the coupling coefficient, given by $M_q = \left[\frac{e^2\hbar\omega_{SO}}{\Omega q}\left(\frac{1}{\varepsilon_{SO}^{\infty} + \varepsilon_{el}(q)} - \frac{1}{\varepsilon_{SO}^{0} + \varepsilon_{el}(q)}\right)\right]^{1/2}$ where $\omega_{SO}$ is the SO phonon energy (Table 1), and $\varepsilon_{SO}^{\infty}$ and $\varepsilon_{SO}^{0}$ are the optical and static dielectric response of the interface, respectively. A large relative difference between $\varepsilon_{SO}^{\infty}$ and $\varepsilon_{SO}^{0}$ results in a large coupling strength. In addition, these parameters can be extracted from optical and capacitance measurements [75]. The term $\varepsilon_{el}(q)$ represents the screening effect of the electron gas on the surface electric field of the substrate SO phonons and like in Froehlich scattering, depends on the temperature as well as the carrier density. At high carrier densities, the difference between the total optical and static dielectric responses becomes relatively small and the coupling coefficient decreases, resulting to less SO phonon scattering.

Figure 2(b) shows the SO phonon-limited mobility $\mu_{SO}$ calculated using Eq. 1 as a function of temperature at $n = 10^{12}$ and $10^{13}$ cm$^{-2}$ for SiO$_2$, Al$_2$O$_3$ and HfO$_2$. We observe that $\mu_{SO}$ is markedly reduced at higher $n$ because of stronger screening of the remote interaction with the SO phonons. Also, $\mu_{SO}$ decreases sharply with increasing temperature because SO phonons are more thermally populated at higher temperatures. Thus, at room temperature, it can be one of the most dominant scattering processes. The SO phonon-limited mobility is particularly low for HfO$_2$ because of its low characteristic phonon frequency and the large difference between its optical and static dielectric response. On the other hand, *h*-BN, which is frequently used as a substrate for MoS$_2$ devices because of its high-quality interface, has a high characteristic SO phonon frequency ($\omega_{TO}$ = 97.4 meV) that leads to a low SO phonon thermal population and considerably less scattering at room temperature.

Surface optical phonon scattering becomes even more significant in dual-gated devices where the MoS$_2$ is sandwiched between an insulating oxide substrate and a top gate oxide. Typically, a thin layer of a high-κ oxide such as HfO$_2$ (κ ≈ 20) is used for the gate oxide to provide better carrier density modulation at lower gate voltages.

However, electrons in the MoS$_2$ experience SO phonon scattering by the phonons in the substrate oxide (e.g. SiO$_2$) and the top gate oxide (e.g. HfO$_2$), resulting in a reduced electron mobility.

**2.3 CI scattering**

Given the atomic thinness of the MoS$_2$ monolayer, the electrons are highly susceptible to external electrostatic influence by long-range sources such as CI, which can be identified with charge centers near the oxide surface and also possibly with sulfur vacancies within the MoS$_2$. CI scattering is known to be a major source of resistance to electron conduction in 2D crystal (e.g. graphene and MoS$_2$) [90, 94, 97] samples and can be invoked to explain the variability of measured mobility values and its discrepancy with theoretical limits predicted from intrinsic electron-phonon scattering rates. Thus, the significance of this scattering mechanism for the interpretation of experimentally extracted mobility values warrants a more in-depth discussion of the underlying physics. The physical picture of CI scattering is further complicated by the phenomenon of charge polarization within the 2D electron gas around the bare CI. The polarized charge around the CI results in the screening of its electrostatic potential and reduces the CI scattering rate. In addition, the screening has a subtle but significant effect on the temperature dependence of the CI-limited mobility. At higher temperatures, the screening effect is weakened and CI scattering becomes stronger, resulting in higher electrical resistance. It was pointed out by *Ong* and *Fischetti*[90] that this can result in the CI-limited electron mobility in monolayer MoS$_2$ having a temperature dependence that is similar to and can be mistaken for the temperature dependence of phonon-limited charge transport.

We assume that the CI originate near the semiconductor-dielectric interface and that, for the purpose of simplicity, they are within the same two-dimensional plane as the electrons. Hence, the screened potential felt by the electron can be expressed as[90]

$$\phi_q^{scr} = \frac{\phi_q}{\varepsilon_{2D}(q,T)}, \quad \phi_q = \frac{e^2}{(\varepsilon_{box}^0 + \varepsilon_0)q}$$

where $\phi_q = \frac{e^2}{(\varepsilon_{box}^0 + \varepsilon_0)q}$ is the bare potential; $\varepsilon_{box}^0$ and $\varepsilon_0$ are in turn the static permittivity of the substrate and vacuum. The screening of the bare CI by the substrate and the free electrons is described by the generalized screening function

$$\varepsilon_{2D}(q,T) = 1 + \frac{2\varepsilon_{el}(q)}{\varepsilon_{box}^0 + \varepsilon_0}, \quad \varepsilon_{el}(q) = -\frac{e^2}{2q}\Pi(q,T,E_F)$$

where $\varepsilon_{el}(q) = -\frac{e^2}{2q}\Pi(q,T,E_F)$ corresponds to the electronic part of the dielectric function. Figure 3(a) shows the representations of the electrostatic potential of the CI for different values of values of carrier densities and substrate oxide types. We see that the effective range of the CI is reduced when the carrier density increases from $10^{12}$ to $10^{13}$ cm$^{-2}$, resulting in stronger charge screening, and when a high-$\kappa$ oxide like HfO$_2$ ($\kappa = 16$) is used in place of SiO$_2$ ($\kappa = 3.9$) for the substrate.

The CI scattering rate is[90]

$$\Gamma_{CI}(E_k) = \frac{n_{CI}}{2\pi\hbar}\int dk\,|\phi_{k-k'}^{scr}|^2(1-\cos\theta_{kk'})\delta(E_k - E_{k'}), \tag{3}$$

where $\theta_{kk'}$ is the scattering angle between the $k$ and $k'$ states and $E_k$ is the energy. The scattering rate is varies linearly with $N_{CI}$, the Coulomb impurity concentration, which has a typical value of ~$10^{12}$ cm$^{-2}$. This implies that the electron mobility is inversely proportional to the impurity concentration. The CI-limited mobility $\mu_{CI}$ can be calculated using Eqs. (1) and (3). The effects of varying the carrier density and the substrate on the CI-limited electron mobility can be seen in Figure 3(b). At $n = 10^{12}$ cm$^{-2}$, the mobility values are significantly lower than the mobility values at $n = 10^{13}$ cm$^{-2}$ because of charge screening. For the same carrier density, the use of a high-$\kappa$ oxide substrate like HfO$_2$ also considerably increases the electron mobility because of more effective dielectric screening. More strikingly, we observe that the electron mobility has significant temperature dependence. At $n = 10^{13}$ cm$^{-2}$, the mobility decreases with rising temperatures, a phenomenon that is commonly associated with electron-phonon scattering but is much more likely to be due to CI scattering in samples with mobility much lower than the theoretical limit of ~410 cm$^2$/Vs [54, 90].

In dual-gated devices, the effect of dielectric screening is more pronounced as the MoS$_2$ is sandwiched between the substrate and the top gate oxide, which changes the effective permittivity in the MoS$_2$. In bare MoS$_2$ on a SiO$_2$ ($\kappa = 3.9$) substrate, the effective permittivity is the average of SiO$_2$ and air ($\kappa = 1.0$), *i.e.*, $\kappa = 2.5$. When the MoS$_2$ is encased by HfO$_2$ ($\kappa = 16$) top gate oxide, the effective permittivity becomes $\kappa = 10$. This 4-fold increase in the permittivity can sharply reduce CI scattering and potentially improve the electron mobility although, as mentioned earlier, the SO phonon scattering from HfO$_2$ can also have a deleterious effect on the mobility. The overall effect of using a top HfO$_2$ gate oxide layer will depend on the relative size of CI scattering to SO phonon scattering. For a low-quality sample with a high concentration of CI, the mobility improvement from reduced CI scattering may dominate the mobility degradation from SO phonon scattering.

**2.4 Atomic defect scattering**

The structural defects have been widely observed in $MoS_2$ and other TMDCs regardless of synthetic methods, including vacancies, dislocations and grain boundaries.[76, 98-103] Compared to graphene[104, 105], higher density of point defects and boundaries is found in $MoS_2$ due to relatively low formation energy.[98, 101, 106] For example, *Jin* et al. investigated the defects in monolayer $MoS_2$ prepared by different methods by annular dark-field scanning TEM (ADF-STEM) (Figure 4a).[107] Through the quantitative analysis of the STEM image, various defect types, such as sulfur vacancies (SV) and antisite defects, can be clearly distinguished. Figure 4b shows that the main type of defects highly depends on sample preparation method. In mechanically exfoliated and CVD samples, which are most widely adopted for making FETs, SV is the dominant point defect type, consistent with most structural studies. [100, 107] However, the reported density of SV has a large variation depending on the samples and characterization techniques.[107-109] For more detailed discussions of atomic defects in TMDCs, we direct the readers to a recent review by *Sun* et al.[110]

Defects can strongly modulate the electronic properties of TMDCs. Since SV are the most dominant defects in $MoS_2$, it is important to understand their influence on the electronic properties. Fig. 4c shows the band structure of $MoS_2$ with SV, calculated by density functional theory (DFT).[98, 111, 112] The presence of SV introduces a deep donor state 0.4-0.6eV below the conduction band edge[100,113]. Real-space mapping of the electron distribution of the mid-gap states indicates that the electrons are localized near the SV, consitent with the large effective mass associated with the weak dispersion.[100] In reality, SV and other defects/impurities in MoS$_2$ form a Gaussian distribution of localized states below the conduction band edge, which is refered to as the mobility edge (Fig. 5a). [100, 112-114]

These localized mid-gap states have important implications in the charge transport of MoS$_2$. When the Fermi energy is below the mobility edge, the charge transport is dominated by Mott variable-range hopping (VRH). The signature of this regime is the temperature scaling of conductivity $\sigma \sim exp[-(\frac{T_1}{T})^{1/3}]$, where the 1/3 in the exponent is due to the 2D nature of MoS$_2$.[115,116] VRH in MoS$_2$ have been observed by many groups, especially in low-mobility samples and at low carrier density.[115-120] *Yu* et al. combined TEM characterizations, DFT calculations and electrical transport measurements to study the hopping transport in monolayer MoS$_2$. The fitting parameters in the VRH are consistent with the measured density of SV, as well as the electron localization length near a SV. Therefore, they concluded that the mid-gap states introduced by SV are mainly responsible for the hopping transport.[100] In addition, SV and other atomic defects can act as centers of short-range scattering. The net effect is to have a constant short-range-scattering-limited mobility $\mu_{SR}$ independent of temperature and carrier density in Eq. 2. as in our early works[76]. However, we find that this term plays secondary roles compared to CI and charge traps, especially near room-temperature where the transistor performance is most relevant. Therefore, for the sake of simplicity, we do not explicitly include this term in this Review.

**2.5 Charge trap and metal-to-insulator transition**

The effects of charge trap on transport were widely investigated and reported in bulk materials, such as silicon [121, 122] and organic materials [123, 124]. TMDCs are no exception. The lattice defects, [100] chemisorption [125], dangling bonds and interface roughness of the dielectric substrate [74, 76] can all act as the sources of traps. *Zhu* et al. carefully measured the capacitance and ac conductance in monolayer CVD MoS$_2$ FETs.[114] By fitting the experimental data, they quantitatively extracted the density and distribution of traps and their time constant as a function of gate voltage (or equivalently, Fermi energy) (Figure 5b). Two types of traps exist, which correspond to mid-gap („M") and band-edge („B") respectively, likely due to different origins. The density of band-edge traps is three orders of magnitude higher than that of mid-gap traps close to the conduction band, suggesting that the former may have stronger influence on charge transport.

Figure 5a draws a cartoon of the density of states (DOS) that explains the different transport regimes in MoS$_2$. Here the mobility edge is the boundary separating the extended states (yellow) and localized states (bule). When the Fermi energy, which is related to carrier density or gate voltage, is above (below) the mobility edge, the electrons propagate as Bloch waves (hopping), showing matellic (insulating) transport behavior. Such trap-induced metal-to-insulator transition (MIT) has been widely reported in many TMDC FETs,[31, 76, 118, 125-128] although electron-electron interaction has also been considered as possible explanation[94]. The MIT transition in TMDC FETs is dominated by extrinsic factor, where the original physical mechanism is transition between band transport and hopping/trap limited transport, which is different with transition introduced by electron-electron interaction in traditional 2DEG.

To quantitatively understand the trap-limited transport in MoS$_2$, we developed a model to account for the reduced carrier density and calculate the effective carrier mobility.[54] Essentially, the charge traps reduce the effective mobility by reducing the density of electrons in the extended states. Assuming that the charge traps have a single Gaussian distribution below the conduction band edge, with a characteristic width $\Delta E_{tr}$. The Fermi energy is determined by the equation

$$n = N_0 \int_0^{+\infty} \frac{dE}{e^{(E-E_F)/k_BT}+1} + \frac{N_{tr}}{\Delta E_{tr}} \int_{-\Delta E_{tr}}^{0} \frac{dE}{e^{(E-E_F)/k_BT}+1}, \quad (4)$$

where $N_0$ is the density of extended states and $N_{tr}$ is the total density of charge traps. The first term on the right hand side is the density of electrons in the extended states ($n_c$), while the second term is the density of trapped electrons. In the simplest picture, the conduction by trapped electrons can be ignored because hopping is a much less efficient process. The effective mobility $\mu_{eff}$ is the fraction of conducting electrons multiplied by the band mobility,

$$\mu_{eff} = \mu_0 \frac{\partial n_c}{\partial n}. \quad (5)$$

We note that this model works better in high carrier density (when the transistor is at ON state), where the density of trapped electrons is an insignificant portion. This further validates the model because most of the field-effect mobility data is derived at ON state. Another natural consequence of the model is that the critical carrier density between band- and trap-induced transport should be close to $N_{tr}$.[76] As we will see in the next section, this is indeed the case for all the analyzed devices.

## 3. Analyzing the mobility data of monolayer MoS$_2$ FETs

Having established the theoretical model, we next analyze the experimental data from literature and understand the underlying limitations in current MoS$_2$ FETs. In our model, we have three key fitting parameters: $N_{CI}$, $N_{tr}$ and $\Delta E_{tr}$. Ideally, the input data should include four-probe field-effect mobility as a function of temperature and carrier density, to exclude contact effects. However, such complete data sets are rather limited, so we try to search for useable data from literatures as much as possible (Table 3). For the two-probe mobility, all the derived microscopic quantities from our fitting represent an upper bound due to the finite effect of contact resistance.

### 3.1. Analyzing a typical MoS$_2$ FET

Figure 6a shows variable-temperature measurements of four-probe conductivity as a function of gate voltage (σ-$V_g$) for a representative back-gated monolayer MoS$_2$ FET on 10nm HfO$_2$/285nm SiO$_2$ substrate. The curves under different temperature show a crossover near $n_0 = C_gV_g \approx 5.3 \times 10^{12}$ cm$^{-2}$, marked by yellow region. For $n < n_0$ (n > $n_0$), σ monotonically increases (decreases) with temperature, indicating insulating (metallic) transport behavior. As discussed in Section 2.5, this is a signature of trap-induced MIT. We further extract the field-effect mobility $\mu = \frac{d\sigma}{C_g dV_g}$ as a function of temperature within each regime in Figure 6a: $n = 3.5 \times 10^{12}$ cm$^{-2}$ (insulating, Figure 6d), $5.6 \times 10^{12}$ cm$^{-2}$ (transition regime, Figure 6c) and $10.6 \times 10^{12}$ cm$^{-2}$ (metallic, Figure 6b). The $\mu$-T relationship also shows distinctive behavior. At high *n*, the mobility shows a similar metallic behavior and decreases monotonically with temperature, reaching ~800cm$^2$/Vs at 20K. At low n, the mobility initially increases as cooled, reaches the highest value of ~90cm$^2$/Vs near 175K, and then decreases rapidly at lower temperature.

We can perform quantitative analysis of the scattering mechanisms by fitting the $\mu$-T relationships under different carrier densities, using the parameters for HfO$_2$ in Table 1. The blue solid lines in Figure 6b-d are the best fitting results with the same fitting parameters $N_{CI}$=0.83×10$^{12}$cm$^{-2}$, $N_{tr}$=4.9×10$^{12}$cm$^{-2}$, and $\Delta E_{tr}$=50.5meV. The green dash lines are the calculated mobility without charge traps. It is obvious that at low carrier density, the localized trap states in the bandgap are mainly responsible for the decrease of mobility at low temperature (Figure 6d). In this regime, most of the electrons are within the trap states, and the transport is dominated by VRH. However, the effects of traps become diminished at high carrier density because they are completely filled. The trap density $N_{tr}$=4.8×10$^{12}$ cm$^{-2}$ is close to the critical carrier density between band- and trap-induced transport, as expected in our model (see discussions in Section 2.5). Furthermore, using the fitting parameters, the CI- and phonon-limited mobility (including both intrinsic and SO) can be quantitatively calculated, as shown by the red and cyan lines in Figure 6b-d. In Figure 6b, a crossover between CI- and phonon-limited mobility at T=223K, indicating that the mobility is limited by phonon (CI) scattering at room (low) temperature. Under intermediate carrier density, transport is affected by CI, phonons and traps collectively.

**3.2 Analyzing the mobility of monolayer MoS$_2$ FETs from literature**

Next we apply our model to available mobility data in the literature. Table 3 lists the fitting parameters for all the devices and the references. In Figure 7a, we select six representative devices fabricated on different substrates for comparison, including $SiO_2$, $HfO_2$ and *h*-BN. One of them has topgate structure, where the model is modified according to the discussions in Section 2. Our model shows excellent fitting results in all the devices.

In Figure 7b, we summarize the distribution of $N_{CI}$ and $N_{tr}$ extracted from the theoretical fitting for all the devices. The different color regions represent the mainstream methods to improve device performance in the literature. We can see that the data points are surprisingly consistent, despite the fact that the devices come from different groups. The figure clearly shows a positive correlation between $N_{CI}$ and $N_{tr}$, indicating that they partly share the same microscopic origin. However, $N_{tr}$ is roughly an order of magnitude higher than $N_{CI}$, probably because many trap states are deep and difficult to be ionized. In backgate devices, direct exfoliation on bare $SiO_2$ without any surface functionalization gives the highest $N_{CI}$ and $N_{tr}$. This is the main reason for the low mobility on the order of 1-10$cm^2$/Vs in early studies. [24, 94, 129] This type of device (the orange data point in Fig. 7b, by fitting the data from Ref. 76, $N_{CI}$=0.91×10$^{12}$cm$^{-2}$, $N_{tr}$=9.3×10$^{12}$cm$^{-2}$) can be used as a reference for comparing different transistor structures. Since then, many approaches have been attempted to improve the performance of monolayer $MoS_2$ transistors such as increasing carrier density by high-κ topgate, lowering defect density by chemical means, and interface engineering. Next we discuss each approach in more depth.

Many topgate MoS$_2$ FETs exhibit superior performance than their backgate counterparts.[94, 130-132] *Kis* et al. used 30nm HfO$_2$ as the topgate dielectrics and reports room-temperature and low-temperature mobility of 63cm$^2$/Vs and 174cm$^2$/Vs respectively.[94] However, the extracted $N_{CI}$ and $N_{tr}$ in topgate devices are much higher as shown in the purple region in Fig. 7b. This is not surprising because the high-κ deposition process inevitably introduces new CI and charge traps. The improved mobility in topgate FETs can be attributed to higher carrier density and dielectric screening effects because they use thin high-κ oxides as the dielectric layer. Due to the large gate capacitance, carrier density can reach 3.6×10$^{13}$cm$^{-2}$,[94] much higher than backgate FETs on 300nm SiO$_2$. Such high carrier density strongly suppresses the effect of charge traps, resulting in a metallic transport behavior. [94] In addition, the CI in topgate FETs can be more effectively screened by the high carrier density and high-κ dielectrics. When these mechanisms outweigh the additional scattering by CI and SO phonons, mobility increase is expected.

The second approach is to reduce the defects in MoS$_2$ by optimized CVD growth [118, 126, 133] and by chemical repairing[76]. Many techniques are used in CVD process to grow high-quality TMDCs, including metal-organic CVD (MOCVD)[31], atmospheric pressure CVD (AP-CVD)[119, 133, 134], optimized precursors[31], and introducing oxygen.[133] Many techniques have produced MoS$_2$ samples with improved FET performances. For example, *Schmidt* et al. used sufur and MoO$_3$ as precursors to grow MoS$_2$ under atmospheric pressure. Without any interface engineering, they realized a high mobility of 45cm$^2$/Vs and 500cm$^2$/Vs at room temperature and low temperature respectively.[119] By using $(C_2H_5)_2S$ and $Mo(CO)_6$ as the precursors in MOCVD, *Kang* et al. reported wafer-scale $MoS_2$ with mobility of 30 $cm^2/Vs$ at room temperature.[31] *Zhang* et al. found that $MoS_2$ quality was greatly improved by introducing a small amount of oxygen into low pressure CVD (LP-CVD). The two-probe mobility of monolayer $MoS_2$ reach $90 cm^2/Vs$ without any additional device optimizations.[133] The SV in $MoS_2$ can also be repaired by supplementing the missing sulfur atoms by thiol molecules.[135] In 1996, *Schulz* et al. demonstrated that ethanethiol ($C_2H_5SH$) has potent chemical activity on defective $MoS_2$ suface.[136] *Lee*'s and *Jung*'s group reported that different organic molecules with sulfydryl group could bond at SV sites and improved the electronic performance.[137, 138] We found that a simple top-side coating of (3-mercaptopropyl)trimethoxysilane (MPS) molecules on exfoliated $MoS_2$ followed by thermal annealing can significantly reduce the SV, resulting in 1.5 fold increase in room-tempeature mobility.[76] Compare to the exfoliated sample without any treatments, these attempts can reduce $N_{CI}$ and $N_{tr}$ by ~10% and ~35%, respectively, as shown by the light-blue region in Fig. 7b. However, we can see that even with reduced defect density, the transistor performance is still moderate and far from phonon-limit.

The last, and perhaps the most effective, approach to improve the mobility is interface engineering. Due to the ultrathin nature of monolayer TMDCs, interface plays key roles in the charge transport. Substrate dangling bonds, surface roughness and absorbates can all contribute to CI and traps. Since the electrons in monolayer TMDCs cannot polarize and screen the electric field in the out-of-plane direction, they are especially susceptible to interfacial CI. The commonly used interface engineering methods are interface passivation, high-κ dielectrics, and h-BN encapsulation. Self-assembled monolayer (SAM) has been widely adopted to passivate and reduce the CI in oxide substrates. [139-142] A huge library of SAM molecules with different end groups can be used, some of which have additional doping effects. It has been shown that octadecyltrimethoxysilane (OTMS)-treated $SiO_2$ substrate can reduce the CI and increase the mobility of graphene to 47,000cm$^2$/Vs at room temperature. [142] *Zhen* et al. investigated the doping effect of different end groups on $MoS_2$ and found that SAM with –$CF_3$ and –$NH_2$ end groups could act as hole and electron donors respectively, due to the charge transfer between SAM and $MoS_2$.[143] *Lou* et al. adopted SAM with sulfydryl group (thiol) to modify the $SiO_2$ substrates. They found that $MoS_2$ on thiol treated substrates showed a higher room temperature mobility of 18cm$^2$/Vs, 6-fold improvement compared to bare $SiO_2$. They attributed the performance improvement to combined effect of interfacial charge transfer, molecular polarities, reduced densities of defects, and suppression of remote phonon scattering.[144] We developed a double-side MPS treatment method to passivate the interface and repair the atomic defects of $MoS_2$ simultaneously. The bottom layer of MPS was grown on $SiO_2$ using a solution SAM process, followed by mechanical exfoliation of $MoS_2$, MPS coating and thermal annealing. [76] Room temperature (Low temperature) mobility of ~80cm$^2$/Vs (~300cm$^2$/Vs) is achieved in double-side treated monolayer $MoS_2$ FETs. Our analysis shows that $N_{CI}$

=$0.71\times10^{12}$cm$^{-2}$ and $N_{tr}$=$5.2\times10^{12}$cm$^{-2}$ for this device, among the lowest within the SAM passivation category.

The motivation of using high-κ substrate (instead of SiO$_2$) is to reduce the CI scattering by dielectric screening. Our group investigated such effect by depositing a thin layer (~10nm) of HfO$_2$ or Al$_2$O$_3$ on SiO$_2$ and comparing the FET performance on different substrates (using double-side MPS treated MoS$_2$ as channel).[75] Compared to topgate, this approach did not introduce extra CI and charge traps. Under high carrier density, we observed increase of mobility with dielectric constant. The model fitting shows that these devices have similar $N_{CI}$ (grey region in Fig. 7b), and the main difference indeed comes from the dielectric screening effect. On HfO$_2$ substrate, the room-temperature mobility of ~150cm$^2$/Vs is, to the best of our knowledge, the highest reported value for monolayer MoS$_2$ FETs so far. However, the downside of using high-κ substrate is the increased SO phonon scattering. Our modeling shows that when $N_{CI}$ is below $0.3\times10^{12}$cm$^{-2}$, using high-κ is no longer advantageous because the transport is switched to SO-phonon-limited regime. The best strategy is therefore to have low $N_{CI}$ and low-κ substrate like *h*-BN.

It is well known that *h*-BN is an ideal substrate for many 2D materials with extremely clean interface [145]. Graphene encapsulated by BN has shown extremely high mobility of 140,000cm$^2$/Vs, electron mean free path of over 15μm, and many exotic quantum phenomena. [146-150] *h*-BN encapsulation is also expected to further decrease $N_{CI}$ and $N_{tr}$ in TMDCs. For MoS$_2$ samples only encapsulated by top *h*-BN, $N_{CI}$ and $N_{tr}$ are already significantly reduced compared to bare MoS$_2$ on SiO$_2$ (red region in Fig. 7b), resulting in a higher mobility of ~60cm$^2$/Vs (~280cm$^2$/Vs) at room temperature (low temperature).[128] Furthermore, through a combination of double-side h-BN encapsulation and graphene contacts, low temperature mobility of over 1000cm$^2$/Vs is achieved for monolayer MoS$_2$. For multi-layer MoS$_2$, the low temperature mobility is much higher ~34,000cm$^2$/Vs. Benefiting from the ultra-high mobility and small Schottky barrier, Shubnikov–de Haas oscillations were observed in MoS$_2$ for the first time.[127] By quantitative fitting, we find that the density of CI and charge traps sharply decreases down to 0.3-0.4×10$^{12}$ cm$^{-2}$ and 4.0-4.6×10$^{12}$ cm$^{-2}$, respectively, which is the cleanest interface reported so far. *Wang* et al. studied monolayer MoS$_2$ device on *h*-BN substrate and found that the critical carrier density between band- and trap-induced transport was ~ 1.0×10$^{13}$cm$^{-2}$, [151] in line with the derived values from our model. Compared to the double-side MPS treated samples on HfO$_2$, the room temperature mobility in double-side *h*-BN encapsulated samples is still lower, probably due to lower carrier density and more defects. We notice that, even for the best MoS$_2$ samples with double-side *h*-BN encapsulation, CI is still three orders of magnitude higher than graphene on *h*-BN(~7×10$^{10}$cm$^{-2}$)[146], which indicates that CI in MoS$_2$ comes partly from the channel material itself, such as surface absorbates and defects. However, the exact microscopic origin of CI and traps is still not clear and is the subject of future investigations. Finally, we emphasize that although BN encapsulation is an elegant way to study the intrinsic limit of MoS$_2$, it is difficult to scale up the layer-by-layer transfer process for applications.

**4. Device modeling of few-layer MoS$_2$ FETs**

A clear tendency in TMDC transistors is that thicker flakes normally exhibit higher mobility, [50, 152-155] mainly owing to the reduced interaction between distance various interfacial scattering sources and the carriers within the channel. For the purpose of optimizing transistor performances, [50] an in-depth understanding of the thickness-dependent mobility as well as its physical origin is strongly desirable. The transport model of monolayer MoS$_2$ discussed in the previous section employs a zero-thickness approximation for the conduction channel, where the carrier distribution is described by a Delta function in the out-of-plane direction. [75, 76, 156, 157] Such an approximation is appropriate for the monolayer channel but is inaccurate for thicker ones. Hence, the transport model should be modified to account for the non-zero thickness of the channel.

**4.1. Transport model for a generalized FET geometry**

In general, all the scattering rates can be simply expressed as $\Gamma_j \sim \left( \dfrac{U_j F_j}{\varepsilon_{2D}} \right)^2$, where $U_j$, $\varepsilon_{2D}$, and $F_j$ are the scattering matrix of the *j-th* type, the 2D electron polarization function, and the device configurative factors, respectively. The difficulty in modeling few-layer TMDC FETs depends on the number of functional layers involved and the structural symmetry of the carrier distribution and dielectric surroundings, which determines the forms of exact scattering matrix, $U_j$ electron polarization function $\varepsilon_{2D}$, and device configurative factors $F_j$ in calculation.[158]

Figure 8 show the device geometry for four typical systems. For conventional bulk silicon transistor (Figure 8a), only two functional layers (a silicon conduction channel and a SiO$_2$ dielectric) are involved. Within the zero-thickness approximation, the number of functional layers is virtually simplified into two for the monolayer graphene and TMDC systems (Figure 8b). For superlattice or dual-gate silicon transistors (Figure 8c), the symmetric carrier distribution within conduction channels and dielectric surroundings around channels can still result in analytical expressions in most cases, in spite of the trilayer structure. Lastly, a few-layer TMDC FET represents a more generalized system that contains three electronically functional layers and lopsided carrier distribution upon applying gate bias (Figure 8d). The concurrence of the increased number of functional layers and the asymmetry of carrier distribution poses a big challenge to transport modeling. No analytical solution is available and thus time-consuming numerical calculation has to be performed.

So far, few efforts have been devoted to the transport modeling of transistors with few-layer TMD channels.[158, 159] *Jena* et al. are the first to include the carrier lopsidedness into the transport modeling and to consider the shift of the centroid of the carrier distribution with increasing gate bias, as shown in Figure 8e.[159] The forms of carrier distribution are obtained by self-consistently solving Poisson equation and Schrödinger equation to account for the quantum effect. In such a numerical scheme, the forms of carrier distribution and the carrier scattering rates have to be recalculated for each gate bias and channel thickness, hence rather time-consuming.

To address this issue, Li et al. adopt the analytical envelope electron wavefunction of classical bulk Si,[160]

$$\phi(z) = \begin{cases} 0, & |z| > t/2 \\ (b^3/2)^{1/2}(z+t/2)e^{-b(z+t/2)/2}, & |z| \leq t/2 \end{cases} \quad (6)$$

while introducing a variational factor $b = kV_g/t + b_{bulk}$ (where $k$ is a tunable coefficient in unit of V$^{-1}$, and $t$ is the channel thickness, and $b_{bulk}$ is the classical bulk value) to describe the dynamic carrier variation with the gate bias and channel thickness in modeling the few-layer TMDC FETs.[158] Such a form, though simple, is able to bridge $t$ in the whole range and correlate $V_g$ quickly, and thus well describes the dependence of carrier distribution on these two factors. It is easily justified that $b \to b_{bulk}(\infty)$ as $t \to \infty\,(0)$, representing the bulk (or the pulse-like) limit. Besides, the adoption of an analytical carrier distribution allows a strict consideration of complicated device factors including the trilayer device structure, dielectric screening effect, and carrier lopsidedness. The exact formulae of $U_j$, $\varepsilon_{2D}$, and $F_j$ for a few-layer FET can be found in reference [155]. Using this model, they successfully explain the dependence of electrical behavior on thickness. It is found that the carrier scattering arising from interfacial CI is mainly responsible for the thickness dependence in TMDC FETs with relatively low interface quality. This conclusion also corroborates with our analysis on monolayers.

**4.2. Dependence of mobility on thickness**

The dependence of mobility on channel thickness has been experimentally studied by several research groups. [20, 127, 128, 152-155, 158-161] Although the absolute mobility values vary broadly, (Figure 9g), a general tendency in the few-layer regime is that thicker channels show higher mobility.

*Li* et al performed in-depth theoretical and experimental studies on this issue. To ensure identical interfacial CI density, MoS$_2$ samples with consecutive numbers of layers on the same substrate are employed (Figure 9a-b). Similar to the model for monolayers, all CI are assumed to distribute at the channel/dielectric interfaces. Hence there are two impurity sources for a FET channel that are located at the top and bottom surfaces, respectively. The variation of interaction distance ($d_{xi}$, where $x$ is top or bottom, $i$ is the number of layer) between the interfacial CI and carriers with the channel is the main origin for the thickness dependence. Figure 9d shows the experimental mobility values at different gate voltages for the series samples with channel thickness from 1 to 5 layers. Evidently, mobility increases monotonically with increasing channel thickness, as a result from the increased $d$ and reduced CI scattering. Such thickness dependence can be well captured by using the device model shown in the last subsection. Figure 9e shows the simulated data by using a same impurity density of $3 \times 10^{12}$ cm$^{-2}$ for both the top and bottom channel interfaces, which quantitatively agree the experimental mobility curves shown in Figure 9d.

Apart from the mobility-$V_g$ characteristics, the model also captures the feature of thickness dependence. Figure 9f compares the thickness dependent experimental mobility data with theoretical calculation. In the pristine and unannealed samples, a total density of interfacial CI (including both channel surfaces) of $6 \times 10^{12}$ cm$^{-2}$ is required to fit the experiment, which suggests that various CI sources such as gaseous absorbates on channel surfaces and dangling bonds at SiO$_2$ surfaces can induce a high density of scattering centers. This analysis lays the foundation of mobility engineering strategies by improving the interface quality, such as absorbate desorption by thermal annealing and channel passivation with superclean BN encapsulation. Figure 9g summarizes the mobility data from different extents of interfacial cleanliness. [20, 127, 128, 154] Generally, the *h*-BN encapsulated channels exhibit higher mobility than those supported by SiO$_2$ substrates due to improved interface cleanliness. It is important to point out that the thickness dependence would become less important as the interface quality is improved and density of interfacial CI becomes very low. Recently, *Hone* and *Duan* et. al., show the monolayer may show higher mobility than few-layer MoS$_2$ in BN encapsulated device structure (blue dots in Figure 9g). In this sense, the thickness dependence is a natural indication of interface cleanliness and device quality.

## 5. Conclusion and perspective

The low mobility in TMDCs is one of the key limiting factors towards their application in electronics. It has become clear through our analysis that many extrinsic factors are at play. In this review, we develop a semi-classical transport model to account for these factors and calculate the carrier mobility in monolayer MoS$_2$ FETs. The model can quantitatively explain the scaling of mobility with temperature and carrier density and, by fitting the experimental data, obtain the density of CI and charge traps. We apply our model to a wide range of FETs in the literature with different substrates and surface treatments. We find that top-gate devices tend to have the highest CI and traps, while BN encapsulation is the most effective method in reducing CI and traps. Interface engineering by thiol chemistry and high-k dielectric screening also proves to be effective in improving device performance. For few-layer MoS$_2$, we introduce an analytical method to capture the lopsided carrier distribution, and quantitatively explain the increasing mobility with the number of layers. It appears that regardless of monolayer or few-layer MoS$_2$, CI has a great influence on the charge transport. By carefully modeling the FET mobility data, we gain deeper insight of the complexity of the charge transport problem, and suggest a clear path towards intrinsic device performance limit. We believe our model can be facilely adopted for other TMDCs, with a set of modified parameters. The recently demonstrated direct integration of TMDC FETs with silicon CMOS calls for attention to TMD/semiconductor interface as well. In BN encapsulated MoS$_2$ samples, quantum oscillations have already been observed at low temperature. The strong spin-orbit coupling in TMDCs should in principle make quantum valleytronic devices possible, if they are "clean" enough.

**Acknowledgements**

This work was supported in part by National Key Basic Research Program of China 2013CBA01604, 2015CB921600; National Natural Science Foundation of China 61325020, 61261160499, 11274154, 61521001; MICM Laboratory Foundation 9140C140105140C14070; a project funded by the Priority Academic Program Development of Jiangsu Higher Education Institutions; "Jiangsu Shuangchuang" program and 'Jiangsu Shuangchuang Team' Program.

Received: ((will be filled in by the editorial staff))
Revised: ((will be filled in by the editorial staff))
Published online: ((will be filled in by the editorial staff))

**Figure 1.** a) Plot of mobilities vs. bandgap for various semiconductors. The color scale represents the spectrum from ultraviolet (UV) to infrared (IR). b) Schematic of U-shape MoS$_2$ tansistor c) Cross-sectional TEM image of U-shape MoS$_2$ transistor. Scale bar is 10nm. (b) and (c) are reproduced with permission from ref. 71, copyright 2016, IEEE

**Figure 2. Phonon-limited carrier mobility in monolayer MoS$_2$.** a) $\mu_{ph}$ as a function of temperature at n=1.0×10$^{12}$cm$^{-2}$ (blue) and n=1.0×10$^{13}$cm$^{-2}$ (red). b) Theory for surface optical phonon limited mobility for different substrates.

**Figure 3. CI-limited carrier mobility in monolayer MoS$_2$.** a) $\mu_{CI}$ as a function of temperature at n=1.0×10$^{12}$cm$^{-2}$ (blue) and n=1.0×10$^{13}$cm$^{-2}$ (red). for different substrates. b) The real space distribution of screened Coulomb potential for a point charge in MoS$_2$ on SiO$_2$ (upper panels) and HfO$_2$ (lower panels) at different carrier densities. Scale bar is 1nm

**Figure 4. Defects characterization and defects Induced localized states** a) Atomic structures of a single-layer MoS$_2$ by STEM. The SVs are highlighted by green circles. b) Statistical histograms of various point defects in PVD, CVD and ME monolayers. (a) and (b) are reproduced with permission from ref. 107, copyright 2015, Nature Publication Group. c) First-principles calculation of band structure for defective monolayer MoS$_2$ with SVs. (c) is reproduced with permission from ref. 112, copyright 2014, Elsevier Ltd.

**Figure 5. Trap-limited charge transport in monolayer MoS$_2$.** a) Schematic of bandstructure with trap states. Reproduced with permission from ref. 100, copyright 2013, Nature Publication Group b) Density and time constant of trap states as a function of gate voltages. The symbols are experimental results extracted from the capacitance $C_{ms}$ and ac conductance $G_p$. The lines are calculated results. (c) is reproduced with permission from ref. 114, copyright 2014, Nature Publication Group.

**Figure 6. Analysis for typical MoS$_2$ transistor on HfO$_2$/SiO$_2$ substrate** a) Four-probe conductivity as a function of $V_g$ for a representative device on HfO$_2$ substrate. b)-d) Field-effect mobility as a function of temperature under n=1.0×10$^{13}$ cm$^{-2}$, 5.6×10$^{12}$ cm$^{-2}$ and 3.5×10$^{12}$ cm$^{-2}$, together with the best theoretical fittings (solid lines,), the calculated CI-limited mobility (red dashed lines), and the calculated phonon-limited mobility (cyan dash dotted line) and calculated mobility ignoring the trap effects (green dash line). Reproduced with permission from ref. 75, copyright 2015, Wiley

**Figure 7. Analysis for MoS$_2$ transistor from literatures** a) Fitting results of representative devices in literatures. Reproduced with permission from ref. 75, copyright 2015, Wiley, ref. 76, copyright 2014, Nature Publication Group, ref. 94, copyright 2014, Nature Publication Group. ref. 118, copyright 2013, AIP, ref. 127,

copyright 2015, Nature Publication Group, ref. 128, copyright 2015, American Chemical Society b) Phase diagram for key parameters in Table 1.

**Figure 8. Transport model for a generalized FET geometry** (a-d) Device geometry for four kinds of typical electronic systems. (a) bulk silicon. (b) graphene. (c) superlattice. (d) general device. (e) Calculated carrier distribution within a $MoS_2$ channel under different gate biases. Large lopsidedness appears at high gate bias as a result of electrostatic equilibrium. Reproduced with permission from ref. 159, copyright 2012, Nature Publication Group,

**Figure 9. Understanding on the dependence of mobility on thickness** (a) Optical image for as-transferred $MoS_2$ flakes with consecutive numbers of layers from 2 to 6 on a same substrate. (b-c) Corresponding FETs with bottom $SiO_2$ as gate dielectric and a schematic diagram. (d) Calculated field-effect mobility versus gate bias. (e) Dependence of mobility on thickness of $MoS_2$ channel over 25 layers. (f) Comparison between experiment and theoretical calculation. (g) A short summary of mobility versus $MoS_2$ channel thickness under different interfacial conditions. Reproduced with permission from ref. 158, copyright 2013, American Chemical Society.

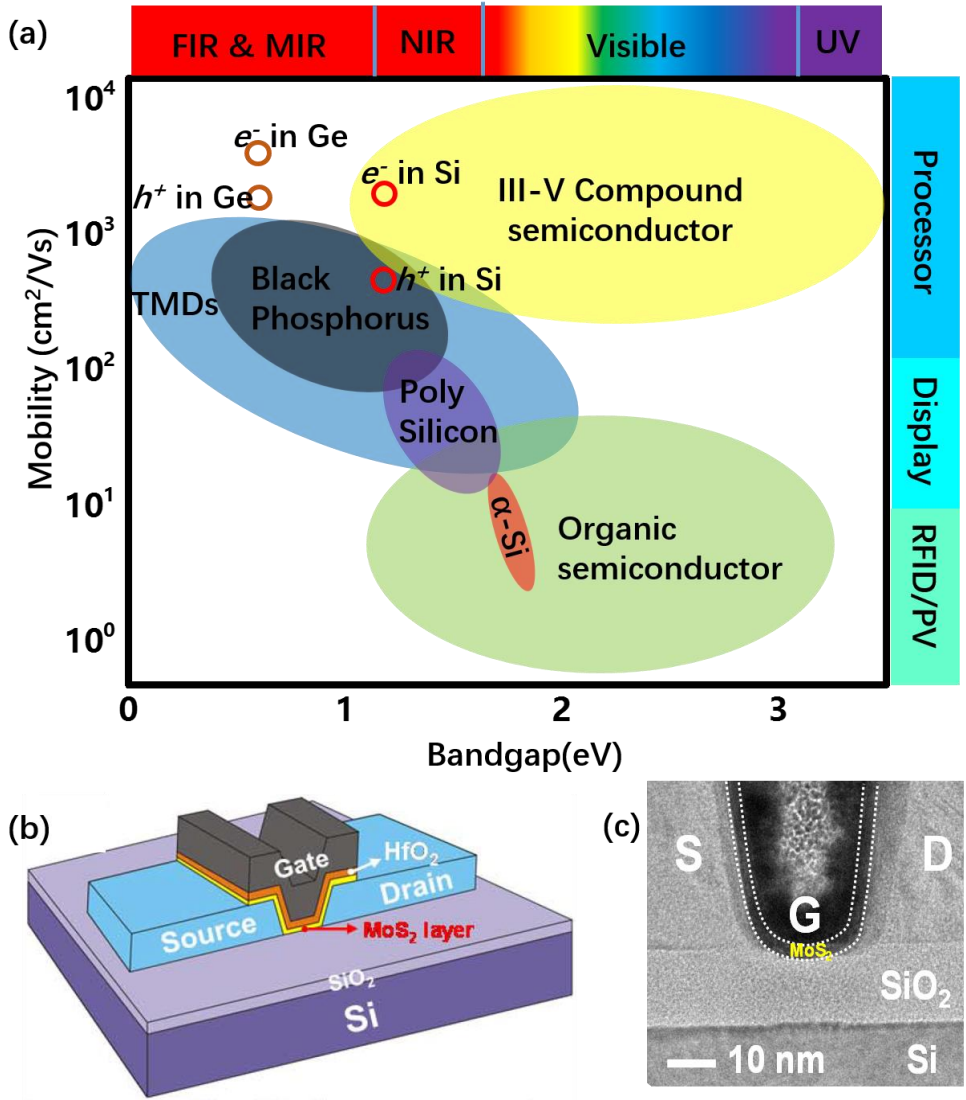

Figure 1

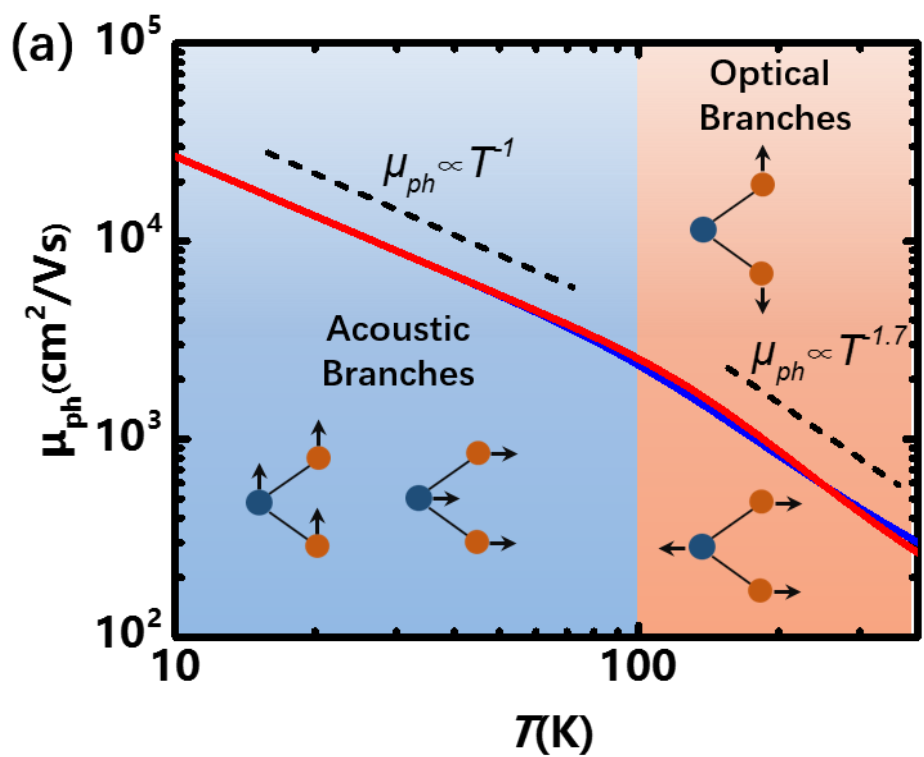

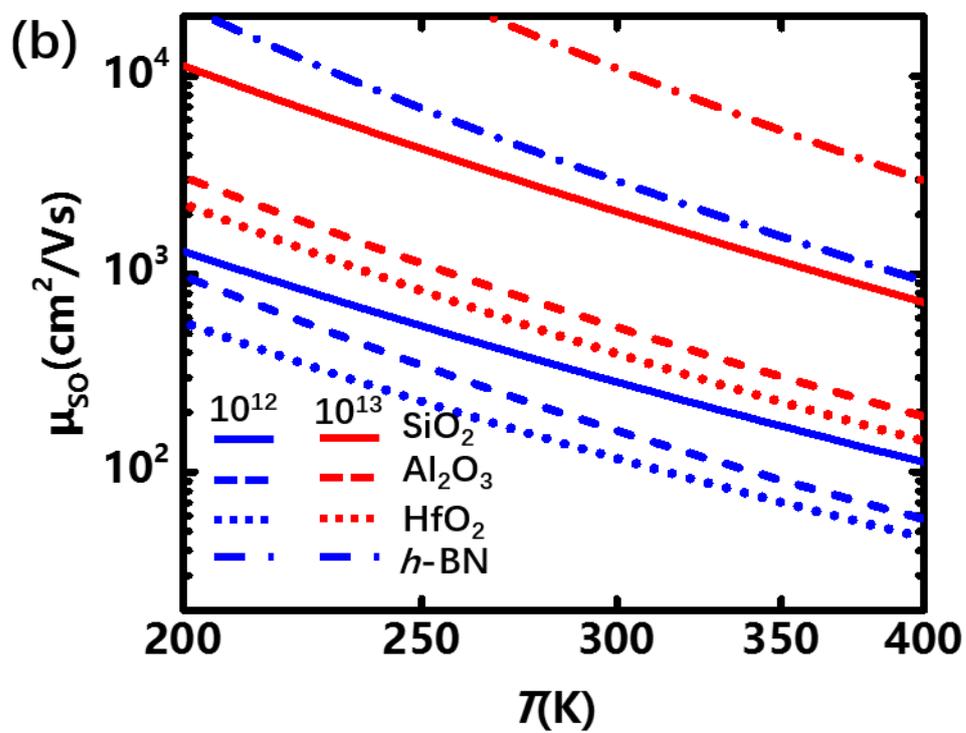

Figure 2

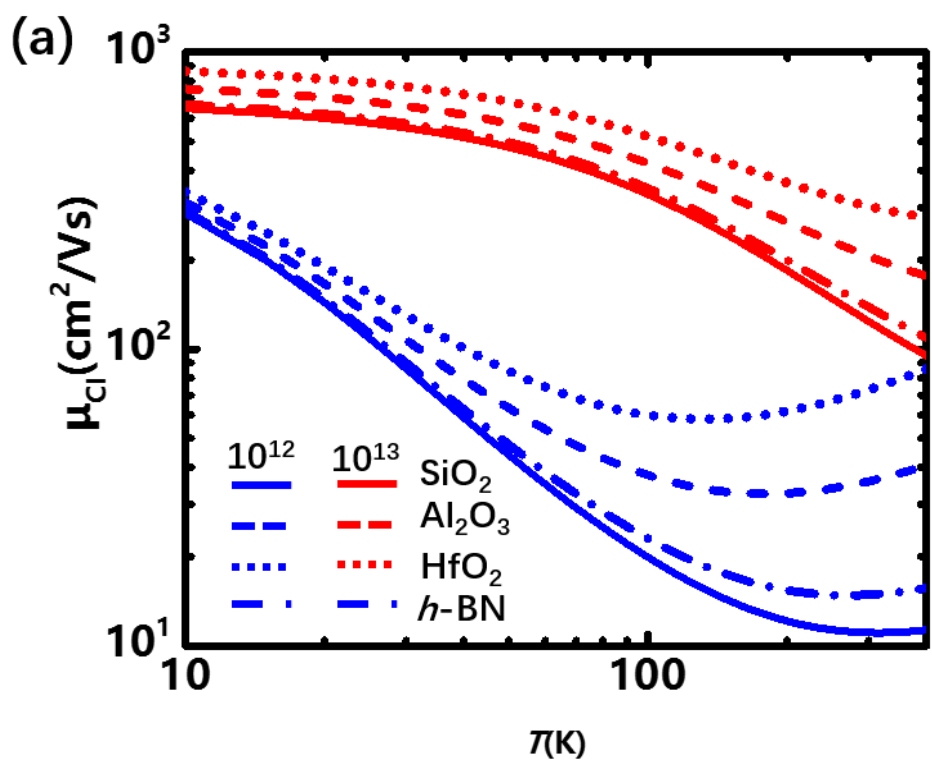
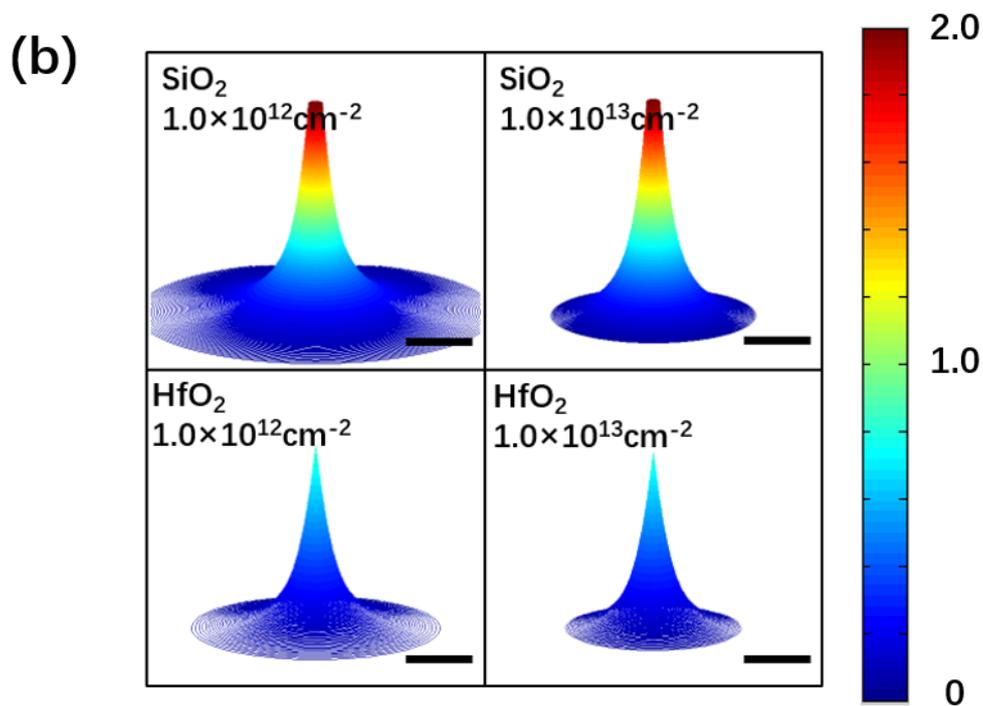

Figure 3

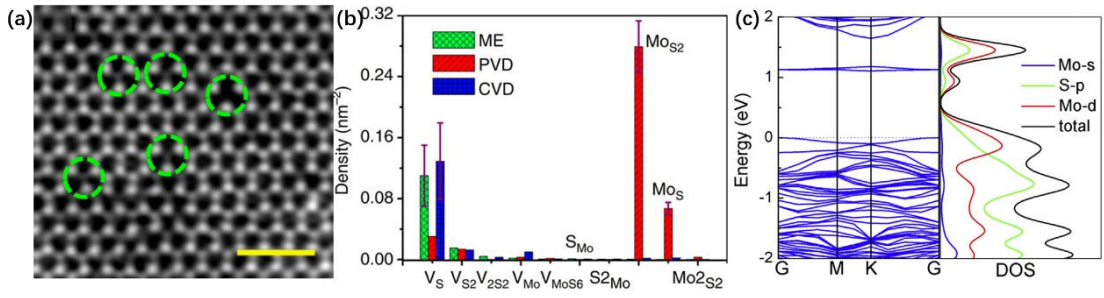

Figure 4

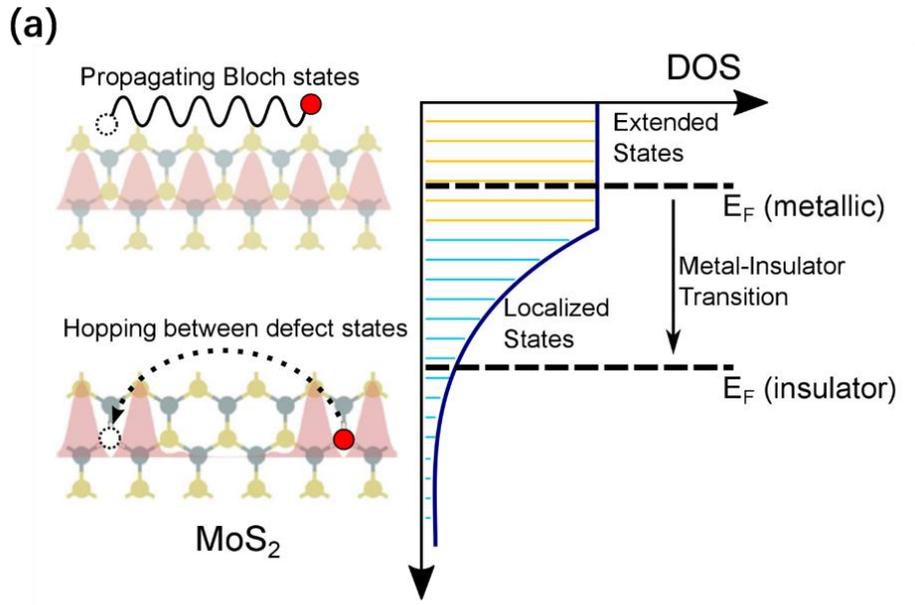

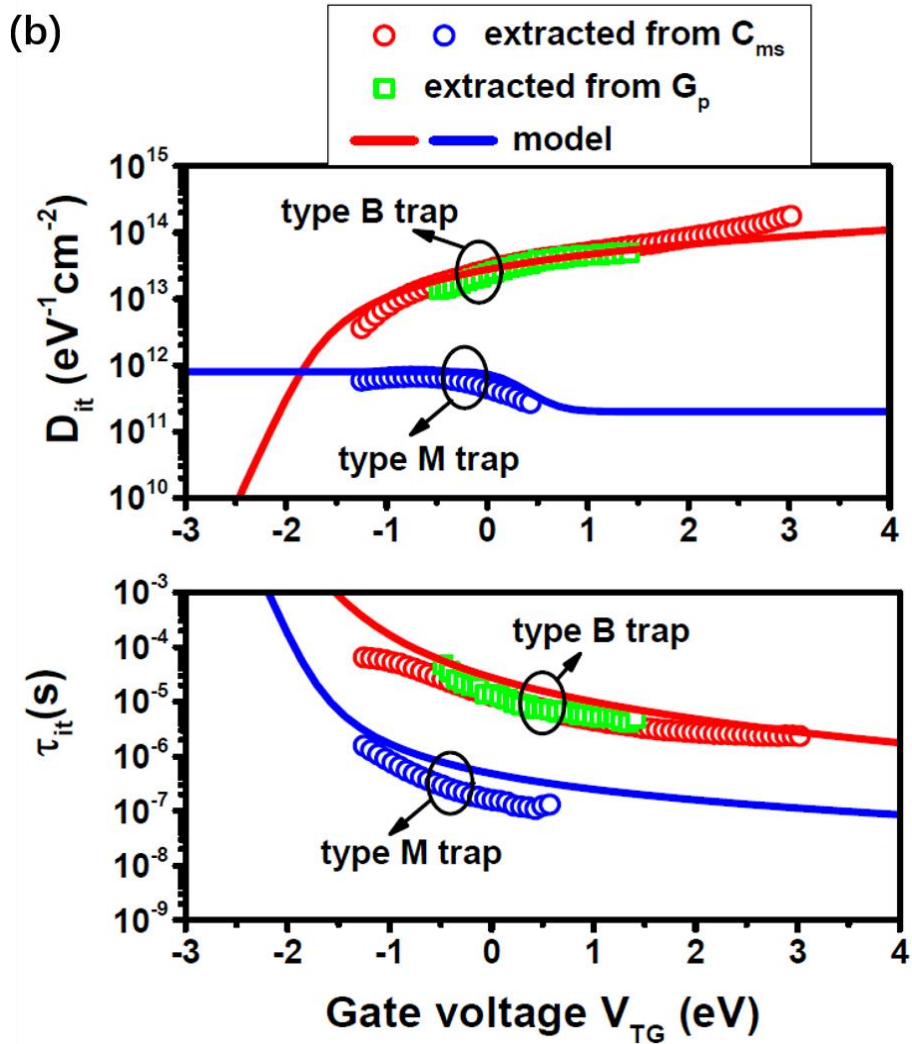

Figure 5

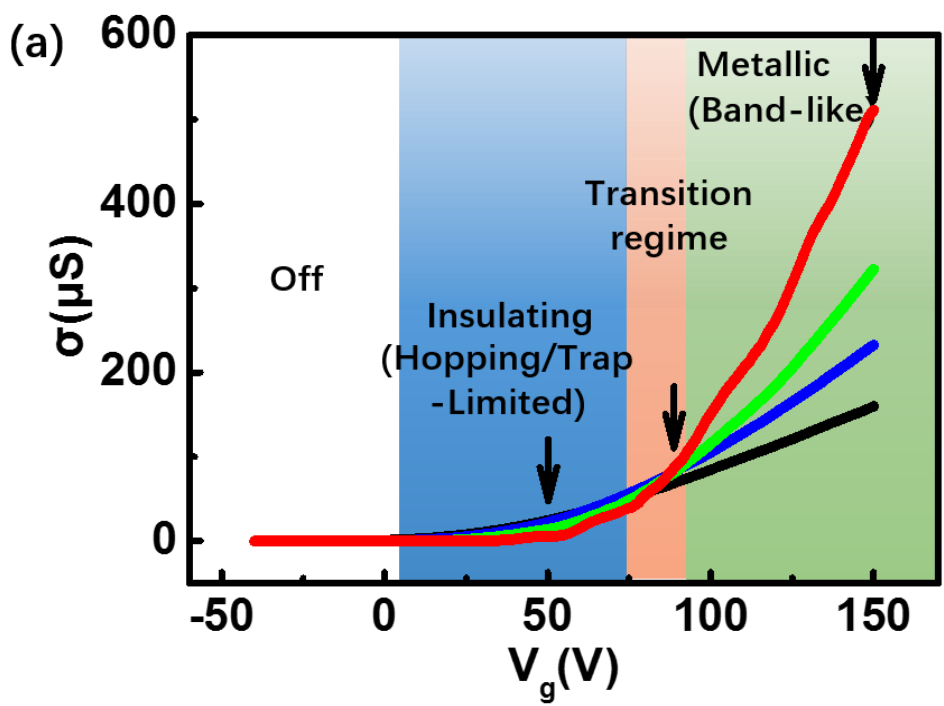
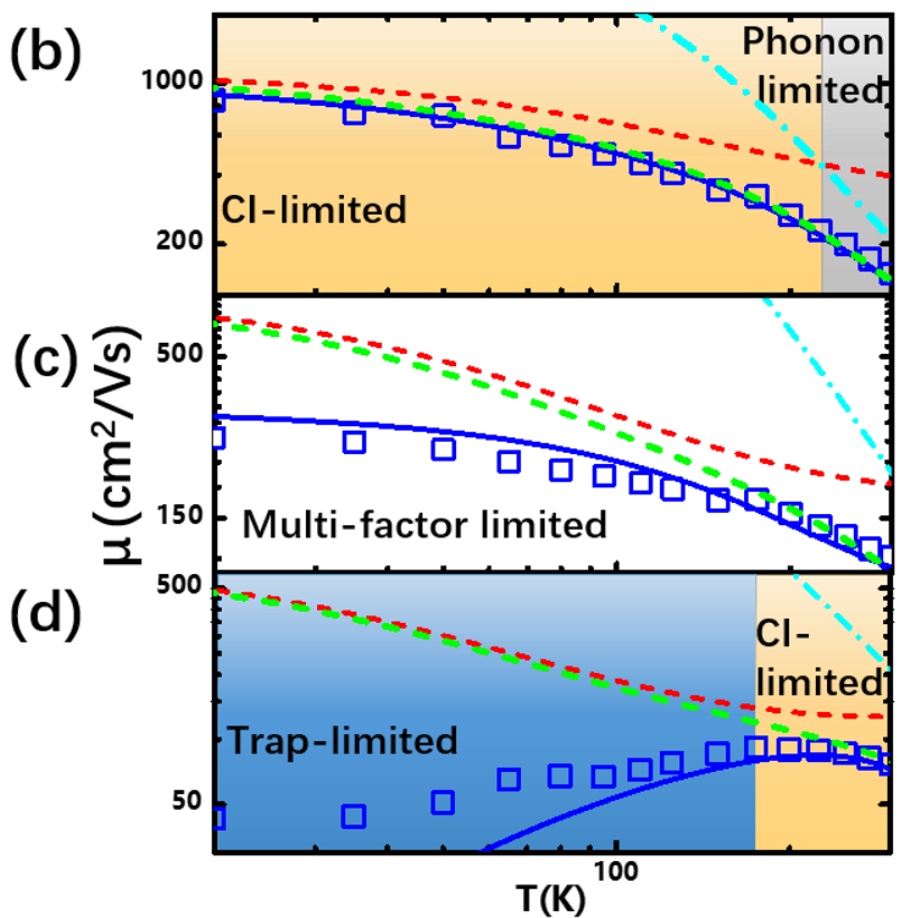

Figure 6

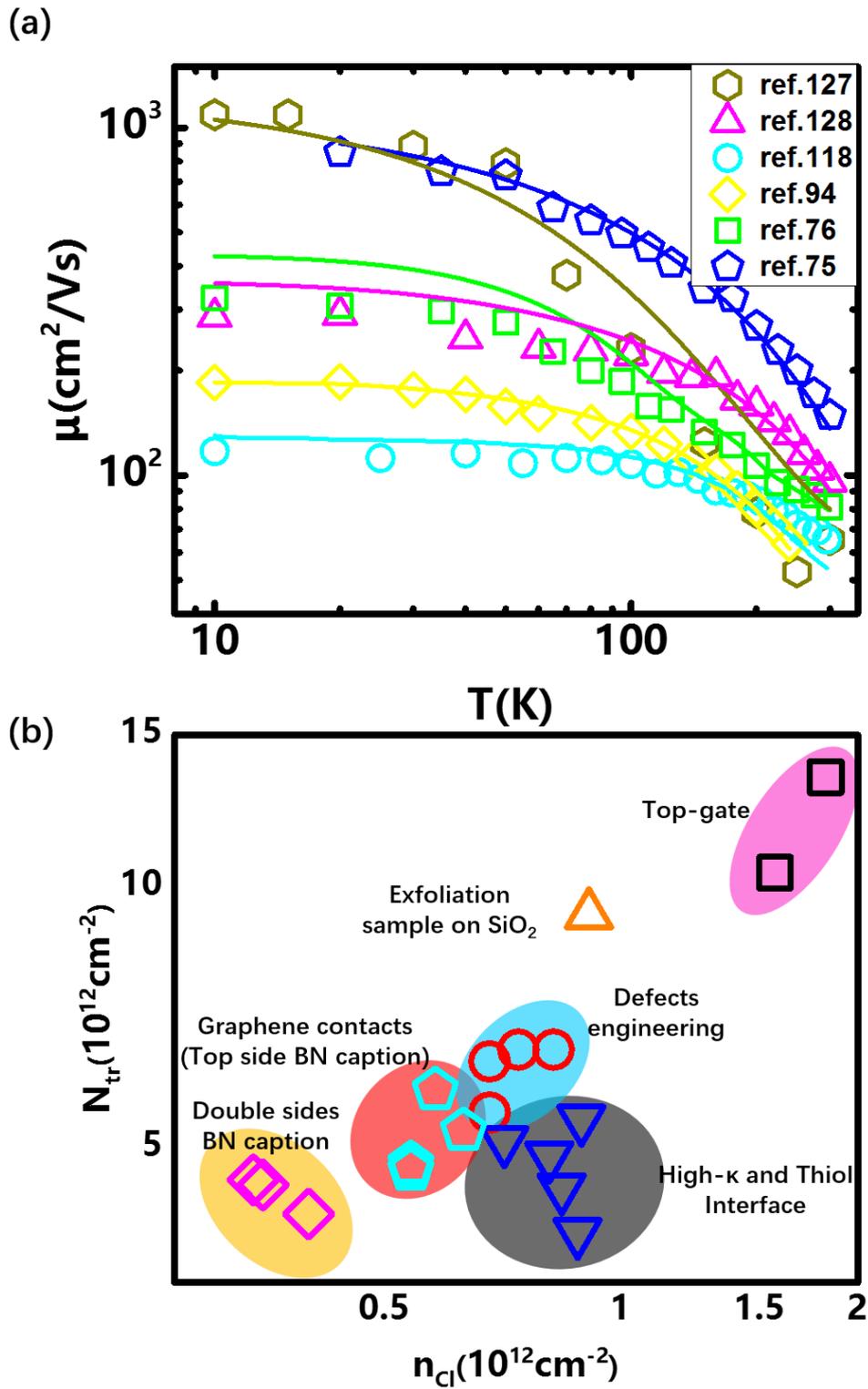

Figure 7

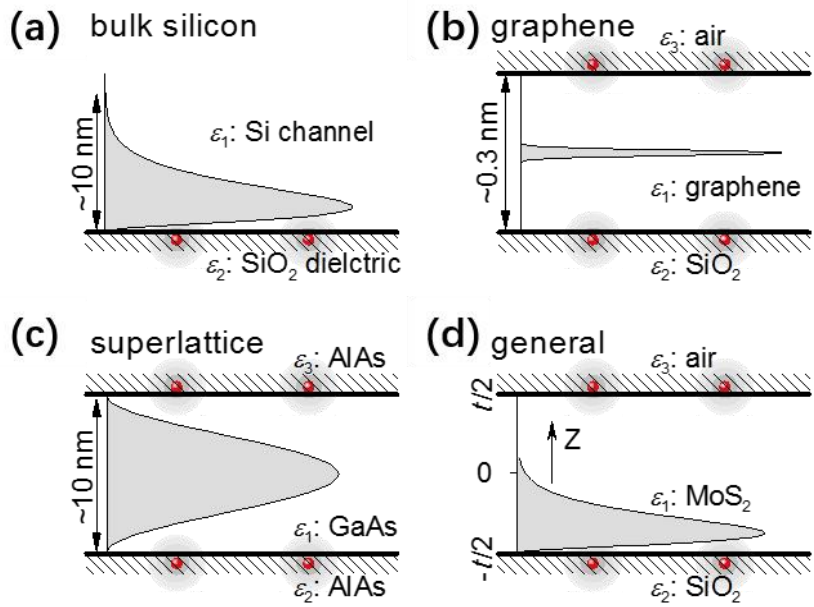
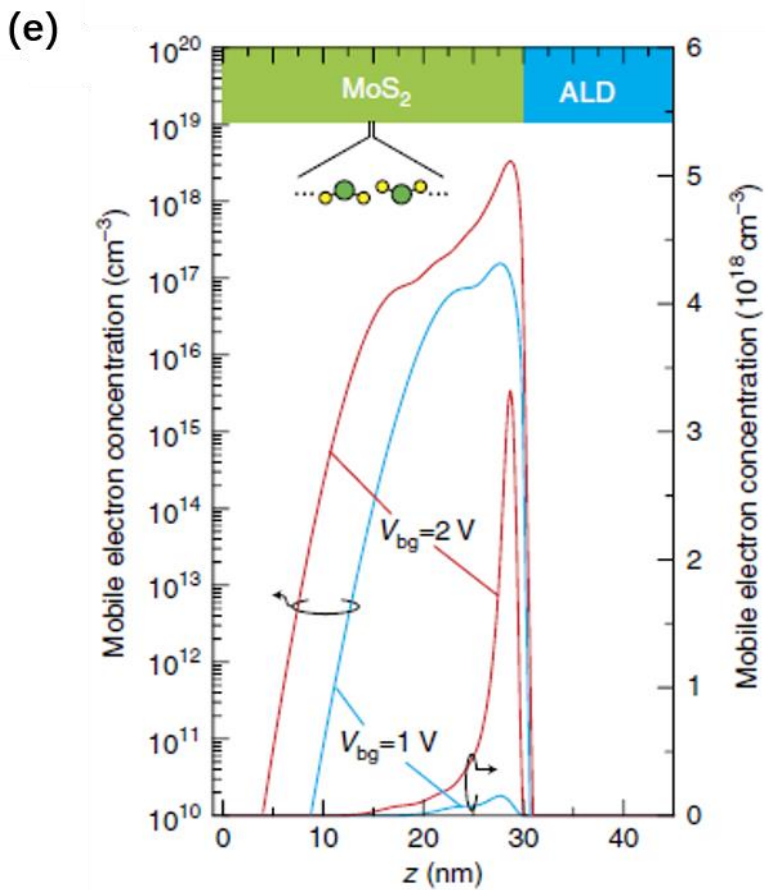

Figure 8

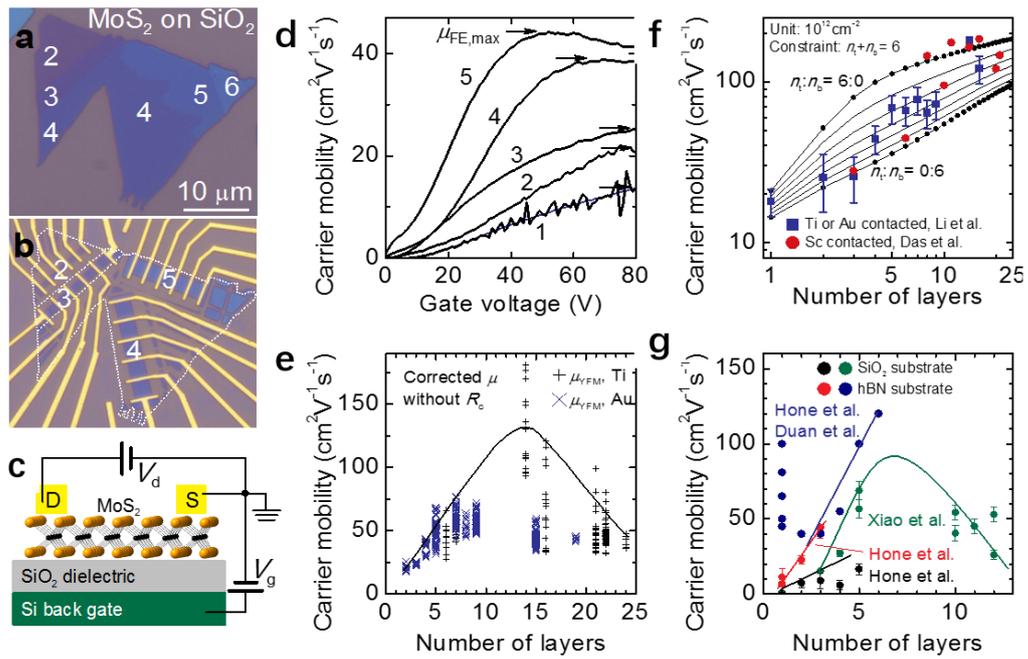

Figure 9

| Parameter | Numerical value |
|---|---|
| $m^*$ | $0.48\,m_0$ |
| $\Xi_{LA}$ ($\Xi_{TA}$) | 2.8 eV (1.6 eV) |
| $c_{LA}$ ($c_{TA}$) | 6700 m/s (4200 m/s) |
| $\rho$ | $3.1\times10^{-7}$ g/cm$^{-2}$ |
| $\sigma$ | $4.41\times10^{-10}$ m |
| $D_{LO}$ ($D_{HP}$) | $2.6\times10^8$ eV/cm ($4.1\times10^8$ eV/cm) |
| $\omega_{LO}$ ($\omega_{HP}$) | 48 meV (50 meV) |
| $\varepsilon_{ion}^0$ ($\varepsilon_{ion}^\infty$) | $7.6\,\varepsilon_0$ ($7.0\,\varepsilon_0$) |

Table 1. Parameters used for the intrinsic acoustic and optical phonon scattering rates.

|  | SiO$_2$ | Al$_2$O$_3$ | HfO$_2$ | Al$_2$O$_3$ (ref.75) | HfO$_2$ (ref.75) |
|---|---|---|---|---|---|
| $\omega_{TO1}$ (meV) | 55.60 | 48.18 | 40.0 | 48.18 | 40.0 |
| $\omega_{TO2}$ (meV) | 138.10 | 71.41 | - | 71.41 | - |
| $\omega_{LO1}$ (meV) | 62.57 | 56.47 | 79.00 | 56.47 | 79.0 |
| $\omega_{LO2}$ (meV) | 153.28 | 120.55 | - | 120.55 | - |
| $\omega_{SO1}$ (meV) | 60.99 | 56.00 | 73.17 | - | - |
| $\omega_{SO2}$ (meV) | 148.87 | 108.00 | - | - | - |
| $\varepsilon_{box}^0$ ($\varepsilon_0$) | 3.90 | 10.00 | 16.50 | 12.53 | 16.00 |
| $\varepsilon_{box}^i$ ($\varepsilon_0$) | 3.05 | 5.80 | - | 7.27 | - |
| $\varepsilon_{box}^\infty$ ($\varepsilon_0$) | 2.50 | 2.56 | 4.23 | 3.20 | 4.10 |

Table 2. Parameters used for the SO phonon scattering rates. The values for SiO$_2$ and Al$_2$O$_3$ (ref.) are taken from Ref. [90]; the values for HfO$_2$ are taken from Ref. [95]. $\varepsilon_{box}^0$ (or κ) for Al$_2$O$_3$ and HfO$_2$ are extracted from capacitance measurements. The other parameters ($\varepsilon_{box}^i$ and $\varepsilon_{box}^\infty$) for Al$_2$O$_3$ and HfO$_2$ are obtained by rescaling the values for Al$_2$O$_3$ and HfO$_2$

| Sample Information | Contacts | $\mu_{RT}$ cm²/Vs | $\mu_{LT}$ cm²/Vs | $N_{CI}$ 10¹²cm⁻² | $N_{tr}$ 10¹²cm⁻² | $\Delta E_{tr}$ meV | Ref. |
|---|---|---|---|---|---|---|---|
| CVD sample (wafer scale), SiO₂ substrate, 2-probe | Ti/Au | 30 | 90 | 0.68 | 5.2 | 57 | 31 |
| Exfoliated, Al₂O₃ substrate, double-side MPS treatment, 4-probe | Ti/Pd | 113 | 465 | 0.84 | 4.5 | 55 | 75 |
| Exfoliated, Al₂O₃ substrate, double-side MPS treatment, 4-probe | Ti/Pd | 101 | 591 | 0.88 | 4.0 | 48 | 75 |
| Exfoliated, HfO₂ substrate, double-side MPS treatment, 4-probe | Ti/Pd | 125 | 540 | 0.89 | 5.4 | 57 | 75 |
| Exfoliated, HfO₂ substrate, double-side MPS treatment, 4-probe | Ti/Pd | 149 | 847 | 0.82 | 4.9 | 50.5 | 75 |
| Exfoliated, SiO₂ substrate, 4-probe | Ti/Pd | 23 | 14 | 0.91 | 9.3 | 82 | 76 |
| Exfoliated, SiO₂ substrate, top-side MPS treatment, 4-probe | Ti/Pd | 34 | 100.5 | 0.82 | 6.5 | 53.5 | 76 |
| Exfoliated, SiO₂ substrate, double-side MPS treatment, 4-probe | Ti/Pd | 81 | 323.8 | 0.71 | 5.1 | 59 | 76 |
| Exfoliated, HfO₂ topgate, Hall measurement | Cr/Au | 66(240K) | 184 | 1.57 | 10.4 | 88.5 | 94 |
| CVD, SiO₂ substrate, 2-probe | Au | 65.3 | 122.8 | 0.74 | 6.5 | 76 | 118 |
| CVD, SiO₂ substrate, 2-probe | Au | 66.4 | 146.7 | 0.68 | 6.3 | 76 | 126 |
| CVD, double-side h-BN encapsulated, Hall measurement | graphene | 65 | 1039 | 0.4 | 4.2 | 42.5 | 127 |
| Exfoliated, SiO₂ substrate, top-side h-BN encapsulated, 2-probe | graphene | 62 | 285 | 0.54 | 4.8 | 50 | 128 |
| Exfoliated, SiO₂ substrate, top-side h-BN encapsulated, 2-probe | graphene | 62 | 269 | 0.54 | 4.72 | 52 | 128 |
| Exfoliated, SiO₂ substrate, 2-probe | graphene | 72 | 136 | 0.58 | 5.2 | 51 | 128 |
| Exfoliated, SiO₂ substrate, 2-probe | graphene | 55 | 124 | 0.62 | 5.25 | 57.5 | 128 |
| Exfoliated, double-side h-BN encapsulated, 2-probe | graphene | 96 | 286 | 0.34 | 4.6 | 49 | 128 |
| Exfoliated, double-side h-BN encapsulated, 2-probe | graphene | 84 | 326 | 0.35 | 4.54 | 42 | 128 |
| CVD, Si₃N₄ substrate, Al₂O₃ topgate, 2-probe | Au | 24 | 58 | 1.82 | 13.4 | 97.5 | 130 |

Table 3. Device Summery and Fitting Parameters

**Keyword**

Two-dimensional; transition-metal dichalcogenides; $MoS_2$; field-effect transistors; mobility; Coulomb impurity.

*Zhihao Yu, Zhun-Yong Ong, Songlin Li, Jian-Bin Xu, Gang Zhang\*, Yong-Wei Zhang, Yi Shi\* and Xinran Wang\**

**Title**

Analyzing the Carrier Mobility in Transition-metal Dichalcogenide $MoS_2$ Field-effect Transistors

**ToC**

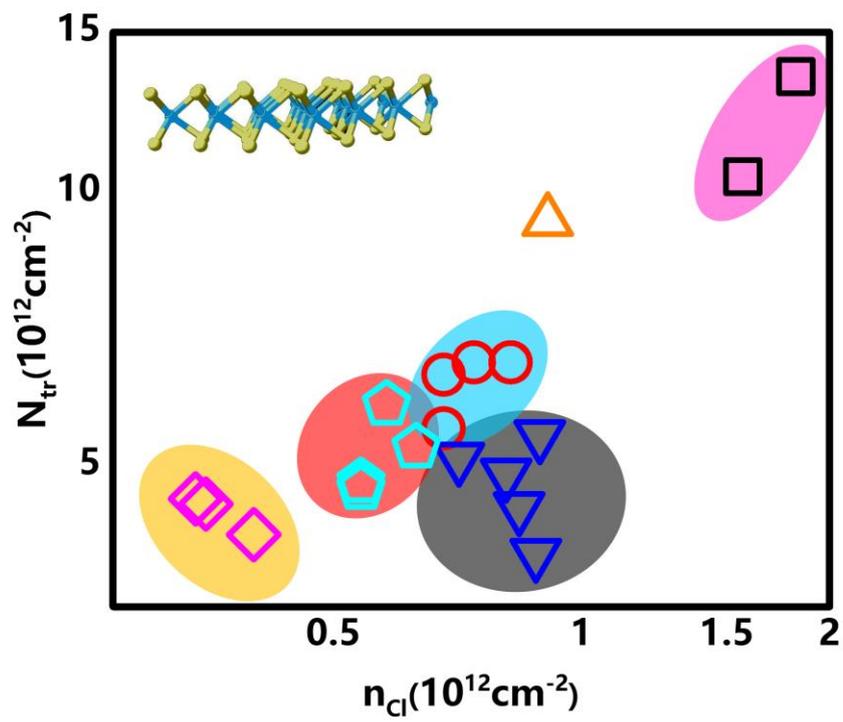

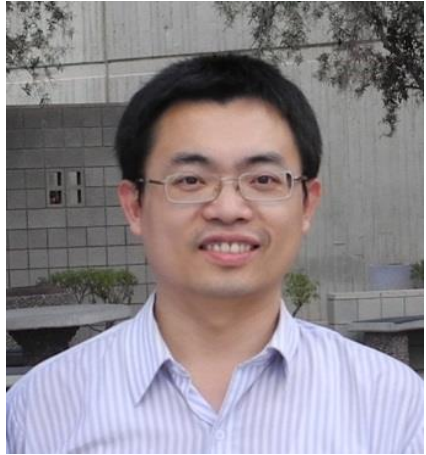

*Gang Zhang is senior scientist and capability group manager in IHPC, Singapore. He received B. Sci /PhD in physics from Tsinghua University in 1998/2002, respectively. His research is focused on the energy transfer and harvesting in nanostructured materials. He has published more than 130 refereed journal papers and 6 book chapters, include 1 in Reviews of Modern Physics. He has edited two books "Nanofabrication and its Application in Renewable Energy"; and "Nanoscale Energy Transport and Harvesting: A Computational Study". He is the editorial board member of Scientific Reports, associate editor of Frontiers in Physics, and technical committee member for IEDM.*

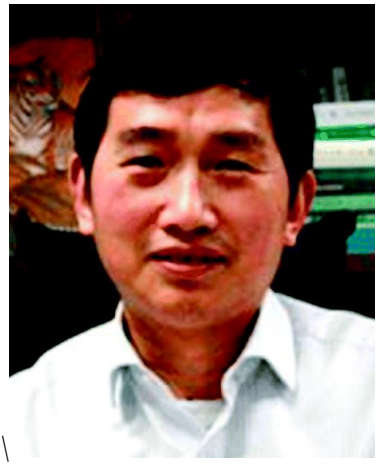

*Yi Shi is full professor and the dean of the school of Electronic Engineering, Nanjing University. He obtained his PhD in the Department of physics, Nanjing University in 1989. He was a visiting scholar at Cambridge University, Tohoku University, and University of California, Berkeley. He has published over 300 peer reviewed papers and was promoted to Changjiang Chair Professor in 2007. His research interests include: nanoelectronics and nano-optoelectronics, and quantum information technology.*

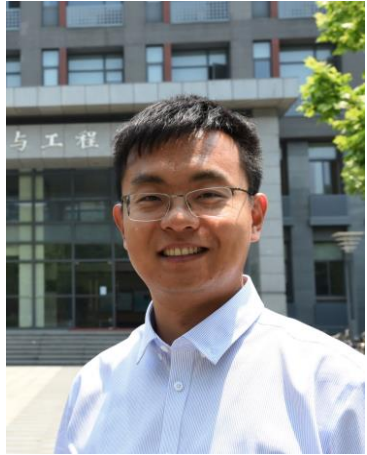

*Xinran Wang received his PhD degree in physics from Stanford University in 2010. Between 2010 and 2011, he was a postdoctoral researcher in Prof. Hongjie Dai group at Stanford University and then in Prof. John Rogers group at University of Illinois at Urbana-Champaign. He joined Nanjing University as a professor of Electronic Science and Engineering in 2011. He is the recipient of the Distinguished Young Scholars award of NSFC (2013), the Chang Jiang Scholar (2014), and the National Youth Metal (2016). Prof. Wang's current research interest includes synthesis, properties and device applications of low-dimensional materials, and flexible electronics.*